\documentclass[preprint, 11pt, amsmath,amssymb,aps,showkeys,showpacs,nofootinbib]{revtex4}

\usepackage{soul}
\usepackage{multirow}
\usepackage{booktabs} 

\usepackage[colorlinks=true, pdfstartview=FitV, linkcolor=blue, citecolor=red, urlcolor=magenta]{hyperref}
\usepackage{graphicx}
\usepackage{latexsym}
\usepackage{amsmath}
\usepackage{amsfonts}
\usepackage{amssymb}
\usepackage{verbatim}
\usepackage{dcolumn}
\usepackage{amssymb}
\usepackage{bm}
\usepackage{float}
\usepackage{physics}
\usepackage[english]{babel}
\usepackage{subfigure}
\usepackage{todonotes}
\usepackage{hyperref}

\graphicspath{{Figures/}}

\begin{document}

\newcommand{\NITK}{
\affiliation{Department of Physics, National Institute of Technology Karnataka, Surathkal  575 025, India}
}

\newcommand{\Shreyasusa}{\affiliation {Department of Oral Health Sciences, School of Dentistry,\\
University of Washington, Seattle, WA 98195, USA.
}}

\newcommand{\naveencroatia}{\affiliation {Theoretical Physics Division, Rudjer Bo\v skovi\'c Institute,\\  Bijeni\v cka c.54,  HR-10002 Zagreb, Croatia}}

\title{Perturbations of Black Holes Surrounded by Anisotropic Matter Field}

\author{Sagar J C }
\email{jcsagar195@gmail.com, sagarjc.226ph028@nitk.edu.in}
\NITK

\author{Karthik R}
\email{raokarthik1198@gmail.com, karthikr.227ph003@nitk.edu.in}
\NITK

\author{Katheek Hegde}
\email{hegde.kartheek@gmail.com}
\NITK

\author{K. M. Ajith}
\email{ajith@nitk.ac.in}
\NITK

\author{Shreyas Punacha}
\email{shreyasp444@gmail.com, shreyas4@uw.edu}
\Shreyasusa

\author{A. Naveena Kumara}
\email{naviphysics@gmail.com, nathith@irb.hr}

\naveencroatia

\begin{abstract}
Our research aims to probe the anisotropic matter field around black holes using black hole perturbation theory. Black holes in the universe are usually surrounded by matter or fields, and it is important to study the perturbation and the characteristic modes of a black hole that coexists with such a matter field. In this study, we focus on a family of black hole solutions to Einstein’s equations that extend the Reissner-Nordström spacetime to include an anisotropic matter field. In addition to mass and charge, this type of black hole possesses additional hair due to the negative radial pressure of the anisotropic matter. We investigate the perturbations of the massless scalar and electromagnetic fields and calculate the quasinormal modes (QNMs). We also study the critical orbits around the black hole and their properties to investigate the connection between the eikonal QNMs, black hole shadow radius, and Lyapunov exponent. Additionally, we analyze the grey-body factors and scattering coefficients using the perturbation results. Our findings indicate that the presence of anisotropic matter fields leads to a splitting in the QNM frequencies compared to the Schwarzschild case. This splitting feature is also reflected in the shadow radius, Lyapunov exponent, and grey-body factors.
\end{abstract}


\maketitle


\section{Introduction}
\label{sec:intro}

The recent breakthroughs in observational astrophysics have significantly advanced our understanding of black hole physics, transitioning it from a predominantly theoretical field to one rich with empirical data. The first direct detections of gravitational waves by the LIGO/Virgo collaborations, originating from the mergers of binary black hole systems~\cite{Abbott:2016blz,TheLIGOScientific:2016src,Abbott:2016nmj}, provided compelling evidence for the existence of stellar-mass black holes and offered unprecedented insights into their dynamics. Simultaneously, the Event Horizon Telescope (EHT) collaboration achieved the groundbreaking imaging of the supermassive black holes at the centers of the galaxies M87 and Sgr A*~\cite{Akiyama:2019bqs,Akiyama:2019cqa,Akiyama:2019fyp}, revealing the shadow cast by these enigmatic objects and illuminating the properties of spacetime in the strong-gravity regime near event horizons. These observational milestones have opened new avenues for probing the universe and necessitate the development of refined theoretical models of black hole physics. Despite the significant progress made in numerical relativity, which allows for fully nonlinear simulations of the Einstein equations without symmetry constraints, perturbative techniques remain indispensable \footnote{See Refs.~\cite{Regge:1957td, Edelstein:1970sk, Zerilli:1970se} for the pioneering works on the perturbation studies of Schwarzschild black hole. Also see the chapter ``Introduction to Regge and Wheeler: `Stability of a Schwarzschild Singularity' '' by Kip Thorne in \cite{Castellani:2019pvh} for a historical account and modern insights on black hole perturbation theory and their relation to quasinormal modes.}. Black hole perturbation theory provides critical insights into the stability of black holes and the emission of gravitational waves, particularly in regimes where full numerical simulations are computationally challenging or infeasible. For instance, perturbative methods have been instrumental in analyzing the late-time behavior of coalescing compact binaries after the formation of an apparent horizon~\cite{Price:1994pm, Abrahams:1994qu, Abrahams:1994xy}.

A central concept in black hole perturbation theory is that of quasinormal modes (QNMs). QNMs are characteristic oscillations of black holes that dominate the gravitational wave signal during the ringdown phase following a perturbation or merger~\cite{Kokkotas:1999bd,Berti:2009kk,Konoplya:2011qq}. These oscillations are described by complex frequencies, where the real part corresponds to the oscillation frequency and the imaginary part represents the damping rate due to the emission of gravitational radiation. Importantly, the QNM frequencies depend solely on the black hole's parameters and are independent of the initial perturbation that excited them~\cite{Chandrasekhar:1975zza, Vishveshwara:1970zz}. As such, they serve as a unique fingerprint of the black hole, enabling the extraction of its properties from gravitational wave observations. Unlike normal modes in conservative systems, QNMs arise in dissipative systems where energy can escape, making the time-evolution operator non-Hermitian. In the context of black holes, the presence of an event horizon leads to intrinsic dissipation, and the associated eigenfunctions are generally non-normalizable and may not form a complete set~\cite{Ching:1998mxl, Nollert:1998ys}. Nevertheless, QNMs play a crucial role in understanding the dynamical response of black holes to perturbations and have analogs in various physical systems, such as leaky resonant cavities and atmospheric waves \footnote{For comprehensive reviews on perturbations and quasinormal modes of black holes and their applications in gravitational wave astronomy, we refer the reader to Refs.~\cite{Kokkotas:1999bd,Berti:2009kk,Konoplya:2011qq,Nollert:1999ji,Nagar:2005ea, Ferrari:2007dd}.}.

In realistic astrophysical environments, black holes are rarely isolated; they are typically embedded in rich surroundings that include matter fields and radiation. The interaction between a black hole and its environment can significantly alter the spacetime geometry and influence observable phenomena such as gravitational waves and shadows cast by them. Understanding this interplay is crucial for interpreting observational data and for making accurate predictions about the behavior of black holes in various astrophysical contexts. While isotropic matter distributions have been extensively studied \cite{Stephani:2003tm, Delgaty:1998uy}, anisotropic matter fields have garnered increasing attention due to their relevance in modeling realistic astrophysical scenarios, such as relativistic stars, self-gravitating systems, and compact stellar objects~(See \cite{Kim:2019hfp} and references therein). Incorporating anisotropic matter fields into black hole solutions introduces new features and deviations from the Schwarzschild and Kerr metrics. For instance, a static, spherically symmetric black hole solution with an anisotropic matter field has been proposed, where the energy-momentum tensor exhibits negative radial pressure~\cite{Cho:2017nhx}. The negative radial pressure allows the anisotropic matter to distribute throughout the entire space from the horizon to infinity, enabling a static configuration with the black hole. This solution generalizes the Reissner-Nordström metric and reduces to it under specific parameter choices. This static solution has been extended to include rotation, leading to a rotating black hole solution with an anisotropic matter field~\cite{Kim:2019hfp}. These solutions have additional parameters related to the density and anisotropy of the surrounding matter field which lead to notable deviations in the black hole's properties and observational signatures. For example, studies have shown that the presence of anisotropic matter fields can alter the shape and size of the black hole shadow~\cite{Badia:2020pnh}, which has potential implications for observations by instruments like the EHT. The influence of anisotropic matter fields on particle collisions has also been studied \cite{AhmedRizwan:2020sza}. Recently, wormholes with anisotropic matter have also been obtained \cite{Kim:2019ojs}. 

Understanding the perturbations and QNMs of black holes surrounded by anisotropic matter fields is essential for several reasons. First, it reveals how anisotropic matter influences the stability and dynamical response of the black hole to perturbations. Second, it affects the gravitational wave signals emitted during events such as mergers or accretion processes, potentially leading to observable differences from standard predictions based on the Schwarzschild or Kerr metric. Third, it contributes to our broader understanding of how matter fields interact with strong gravitational fields in general relativity, which is important for exploring alternative theories of gravity and the nature of dark matter.

In this paper, we study the perturbations and quasinormal modes of static black holes immersed in anisotropic matter fields. We focus on scalar and electromagnetic perturbations and study how the QNM frequencies are modified due to the anisotropic matter. We employ perturbation theory and utilize semi-analytical method to compute the QNM frequencies, considering various values of the anisotropy parameters. Additionally, we explore the connection between the QNMs and the properties of unstable photon orbits, such as the shadow radius and the Lyapunov exponent, which characterize the instability timescale of these orbits. We also analyze the scattering of waves in the black hole spacetime and compute the grey-body factors, which describe the modification of the radiation spectrum due to the potential barrier surrounding the black hole.

The paper is organized as follows. In Sec.~\ref{sec1}, we review the black hole solution surrounded by an anisotropic matter field and discuss its properties and parameter ranges. In Sec.~\ref{sec2}, we derive the perturbation equations for scalar and electromagnetic fields in this background. In Sec.~\ref{sec3}, we compute the quasinormal modes using higher order WKB method and analyze the effects of the anisotropic matter field on the QNM spectrum. In Sec.~\ref{sec4}, we examine the critical orbits around the black hole and their connection to the QNMs, including the shadow radius and Lyapunov exponent. In Sec.~\ref{sec5}, we study the scattering of waves and compute the grey-body factors, discussing their dependence on the anisotropy parameters. Finally, in Sec.~\ref{sec_disc}, we summarize our results and discuss their implications. In addition, Appendix~\ref{errapp} provides details of the error analysis, and Appendix~\ref{gwapp} includes comments on gravitational perturbations.

\section{Black Hole Surrounded by an Anisotropic Matter Field}\label{sec1}

In this section, we present the solution of a black hole surrounded by an anisotropic matter field. The action leading to the field equations corresponding to such black hole solutions is given by \cite{Kim:2019hfp},
\begin{equation}
\mathcal{I}= \frac{1}{16\pi G} \int_{\mathcal M} \sqrt{-g} \, d^4 x \left[(R-F_{\mu\nu}F^{\mu\nu}) +{\cal L}_m \right] +\mathcal{I}_{S} \, ,  \label{action}
\end{equation}
where \( R \) is the Ricci scalar of the spacetime manifold \( \mathcal{M} \), \( F_{\mu \nu} \) is the electromagnetic field tensor, and \( \mathcal{L}_m \) is the Lagrangian density corresponding to the effective anisotropic matter fields. The term corresponding to the anisotropic matter field can result from an extra \( U(1) \) field or other diverse dark matter forms. Without loss of generality, we set \( G=1 \). In the above action, \( \mathcal{I}_{S} \) is the Gibbons-Hawking boundary term, which is essential to make the variational principle well-defined \cite{Gibbons:1976ue, Hawking:1995ap}. The variation of the action \eqref{action} leads to the Einstein equations,
\begin{equation}
G_{\mu\nu}=R_{\mu\nu}-\frac{1}{2} g_{\mu\nu}R=8\pi T_{\mu\nu} \,, \label{einsteineq}
\end{equation}
and the Maxwell equations,
\begin{equation}
\nabla_{\mu}F^{\mu\nu} = \frac{1}{\sqrt{-g}} \, \partial_{\mu}(\sqrt{-g}F^{\mu\nu}) =0 \,. \label{maxwelleq}
\end{equation}
The energy-momentum tensor in \eqref{einsteineq} is sourced by the Maxwell field and the anisotropic matter field, which is given by,
\begin{equation}
T_{\mu\nu}=\frac{1}{4\pi}\left(F_{\mu\alpha}F_{\nu}^{\alpha}-\frac{1}{4}g_{\mu\nu}F_{\alpha\beta}F^{\alpha\beta}\right)-2\frac{{\partial \cal L}_m }{\partial g^{\mu\nu}}+ g_{\mu\nu} {\cal L}_m.
\end{equation}

Starting from a spherically symmetric and static space-time metric ansatz, the field equations lead to the black hole solution described by the following metric \cite{Cho:2017nhx, Kiselev:2002dx},
\begin{equation} 
ds^2=  - f(r) \, dt^2 +  f(r)^{-1} \, dr^2 + r^2 \, d\theta^2 + r^2 \sin^2(\theta) \, d\varphi^2  \,, \label{background_metric}
\end{equation}
with the metric function
\begin{equation}
f(r)=1-\frac{2M}{r} +\frac{Q^2}{r^2} - \frac{K}{r^{2w}},   
\label{metric_function}
\end{equation}
where \( M \) is the ADM mass and \( Q \) is the electric charge of the black hole. The parameters \( K \) and \( w \) control the density and anisotropy of the fluid surrounding the black hole, respectively \cite{Kim:2019hfp, Cho:2017nhx, Kiselev:2002dx}.

The energy-momentum tensor for the anisotropic matter field is \( T^{\nu}_{\mu} = \mathrm{diag}(-\varepsilon, p_r, p_{\theta}, p_{\varphi}) \), where \( p_r(r) = -\varepsilon(r) \) and \( p_{\theta}(r) = p_{\varphi}(r) = w \, \varepsilon(r) \) \cite{Cho:2017nhx}. The negative radial pressure allows the anisotropic matter to distribute throughout the entire space from the horizon to infinity. Therefore, the black hole can be in a static configuration with the anisotropic matter field. The energy density is given by,
\begin{equation}
\varepsilon(r) =\frac{Q^2}{8\pi r^{4}}+ \frac{r^{2w}_o}{8\pi r^{2w+2}} \,,
\end{equation}
where \( r_o \) is a charge-like quantity with the dimension of length, defined by \( r^{2w}_o = (1-2w)K \). However, it should be noted that \( w=1 \) in the neutral \( (Q=0) \) anisotropic matter field solution represents the Reissner–Nordström (RN) like geometry with charge \( r_0 \). Hence, we can say that the RN geometry is a special case of the neutral anisotropic matter field solution. This also emphasizes that the energy around the charged black hole is anisotropic.

For static solutions, the constraints on the parameters \( w \) and \( K \) from the positive energy conditions are given by \( r_0^{2w} = (1-2w)K \geq 0 \) for the neutral case, and \( Q^2 + r_0^{2w} r^{2(1-w)} \geq 0 \) for the charged case \cite{Kim:2019hfp}. Specific combinations of \( w \) and \( K \) correspond to solutions that appear in different spacetime structures. For example, when \( w=0 \), the fluid explains the flat rotation curves of galaxies \cite{Zwicky:1933gu, Rubin:1970zza}. If \( w=1 \), the matter field describes an extra \( U(1) \), which has the same global behaviors of density and pressure as the Maxwell field.

The spacetime metric \eqref{background_metric} reduces to the Reissner-Nordström black hole when \( K=0 \), and to the Schwarzschild solution when \( K=0 \) and \( Q=0 \). When \( K<0 \) and \( w=1 \), it corresponds to the Reissner-Nordström black hole with a constant scalar hair \cite{Zou:2019ays}. The asymptotic flatness of the black hole spacetime is satisfied only for \( w>0 \). However, for \( 0 \leq w \leq 1/2 \), the energy density is not sufficiently localized such that the total energy diverges \cite{Kim:2019hfp}. Therefore, the choice for \( w \) to have asymptotically flat solutions with finite total energy is \( w>1/2 \) (an additional analysis for the \( w=1/2 \) case is given in \cite{Cho:2017nhx}). However, any positive or negative value is allowed for \( K \) (provided the black hole solution exists), which represents diverse matters surrounding the black hole \cite{Lee:2021sws}. In our analysis, we set \( Q=0 \) and focus on the neutral case only (and we choose $M=1$ wherever required).

\begin{figure}[t]
    \centering
    \includegraphics[width=\textwidth]{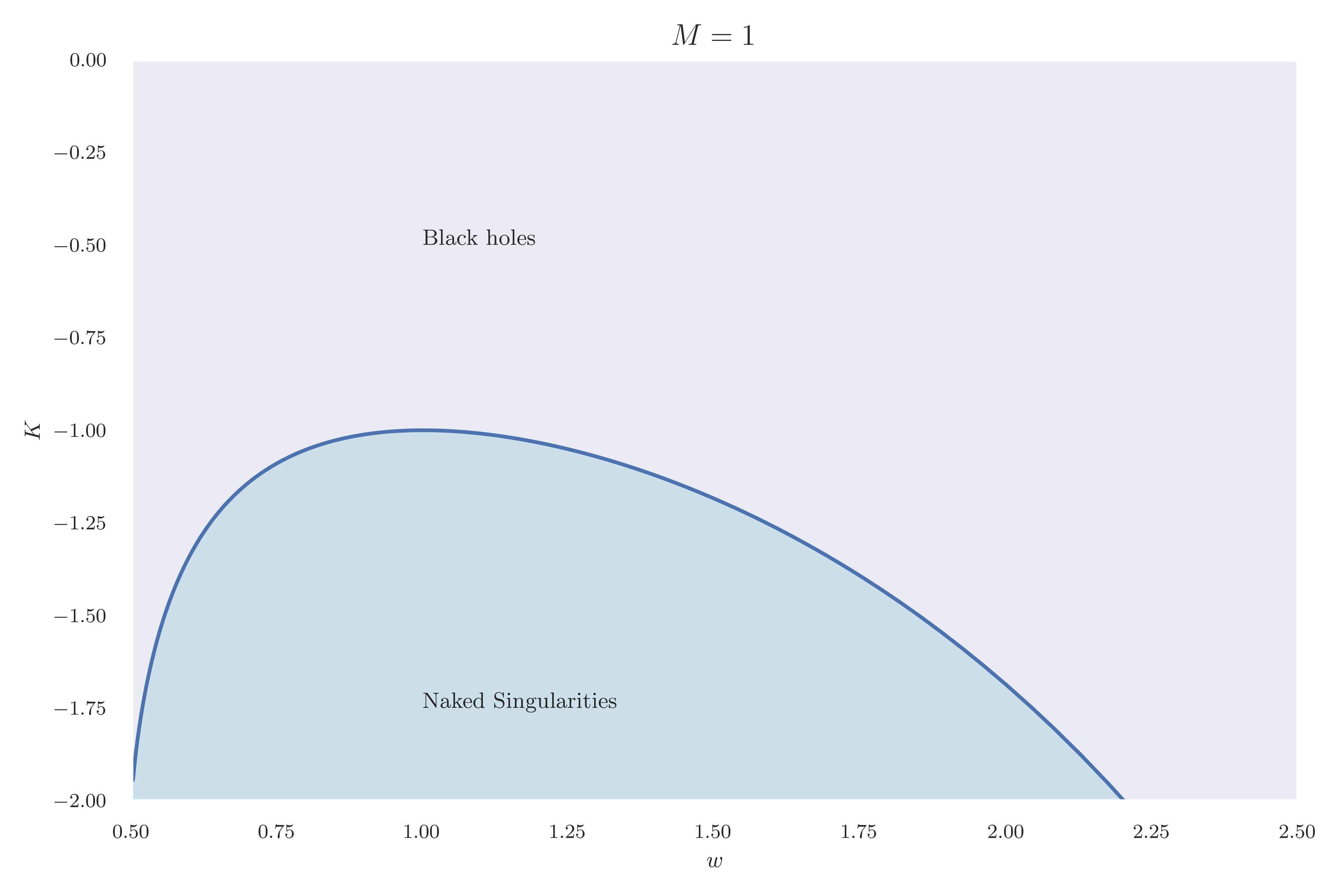}
    \caption{The parameter space of \( w \) and \( K \) showing regions corresponding to black hole and naked singularity solutions. The solid (blue) curve represents the boundary where the event horizon vanishes.}
\label{fig:fig0}
\end{figure}

The metric becomes singular where \( f(r) = 0 \), with the largest root marking the event horizon of the black hole. The disappearance of the event horizon occurs when the equations \( f(r) = 0 \) and \( f'(r) = 0 \) are satisfied simultaneously, allowing us to map out regions in the parameter space of \( (w, K) \) where black hole solutions exist. Specifically, we have the following equations,
\begin{equation}
\begin{split}
f(r) = 0 &\implies r^2 - 2Mr - Kr^{2(1-w)} = 0, \\
f'(r) = 0 &\implies 2(r - M) - 2(1-w)Kr^{1-2w} = 0.
\end{split}
\end{equation}
Because it is not always possible to solve these equations analytically, a parametric solution is often obtained, as outlined in \citep{Badia:2020pnh}. By expressing the first equation for \( K \) and substituting this result into the second equation, we find,
\begin{equation}
w = \frac{M}{2M - r}.
\end{equation}
Substituting this into the initial equation, we obtain
\begin{equation}
K = r^{\frac{r}{2M - r}}(r - 2M).
\end{equation}
The solutions are illustrated in Fig. \ref{fig:fig0} for a fixed \( M \). The solid line distinguishes regions where black hole solutions exist from those leading to naked singularities. 

\section{Black Hole Perturbation}\label{sec2}

Since the black hole is an open system, after a small perturbation, it eventually relaxes to an equilibrium state by losing energy through emitting radiation, depending on the nature of the underlying perturbations. Conventionally, black hole perturbation studies are carried out in two different ways. (i) Perturbation of a test field in the black hole background: Here, the perturbation of external fields in the black hole spacetime is studied without considering the effects of back-reaction. Under this assumption, the perturbation of fields is described by the covariant equation of motion of the corresponding field. (ii) Black hole metric perturbation: This is the \emph{real} gravitational perturbation in which the evolution equation can be obtained by linearizing the Einstein equations. The gravitational radiation emitted in this type of perturbation is much stronger than that from external field perturbations.

\subsection{Scalar Field Perturbation}

First, we consider the massless scalar field perturbation. It is described by the equation of field propagation in the black hole spacetime, which is the Klein-Gordon equation,
\begin{equation}
     \nabla_\mu\nabla^\mu \Psi (t, \vec{r}) = \frac{1}{\sqrt{-g}}\partial_\mu\left( \sqrt{-g}\,g^{\mu\nu}\partial_\nu\Psi (t, \vec{r}) \right) = 0.
    \label{KG_eqn}
\end{equation}
In the background of a black hole with an anisotropic matter field described by the spacetime metric \eqref{background_metric}, the scalar field equation takes the form,
\begin{equation}
    \left[\frac{\partial^{2} \Psi}{\partial r^{2}} + \frac{\partial \Psi}{\partial r}\left( \frac{\partial }{\partial r} + \frac{2}{r}\right)\right] f(r) - \frac{1}{f(r)}\frac{\partial^{2} \Psi}{\partial t^{2}} + \frac{1}{r^{2}}\left[ \frac{\partial^{2} \Psi }{\partial \theta^{2}} + \frac{\cos \theta }{\sin(\theta) } \frac{\partial \Psi}{\partial \theta} + \frac{1 }{\sin^{2} \theta } \frac{\partial^{2} \Psi}{\partial \varphi^{2}}\right]=0.
    \label{KG_eqn_2}
\end{equation}

Due to the spherical symmetry of the spacetime, the variables can be separated by splitting the scalar field \( \Psi(t,r,\theta,\varphi) \) into radial and angular parts,
\begin{equation}
    \Psi(t,r,\theta,\varphi) = \sum_{\ell,m} \frac{\psi(t,r)}{r}\, Y_\ell^m (\theta,\varphi).
    \label{var_sep_eqn}
\end{equation}
Here, \( Y_\ell^m (\theta,\varphi) \equiv Y_\ell^m \) are the spherical harmonics, which are the eigenfunctions of the Laplace-Beltrami operator \( \Delta \), i.e., \( \Delta Y_\ell^m = - \ell (\ell+1)Y_\ell^m \). On substituting Eq.~(\ref{var_sep_eqn}) into the Klein-Gordon Eq.~(\ref{KG_eqn_2}), we obtain the radial part of the perturbation equation as,
\begin{equation} \label{sc_radial}
    \frac{\partial^{2}\psi}{\partial r^{2}}  + \frac{f'(r)}{f(r)}  \frac{\partial \psi}{\partial r}   - \frac{1}{f(r)^2} \frac{\partial^{2} \psi}{\partial t^{2}} - \left(\frac{\ell (\ell+1)}{r^2} + \frac{f'(r)}{r} \right)\frac{\psi}{f(r)}  = 0.
\end{equation}

The above equation can be written as a Schrödinger-like wave equation using the coordinate transformation,
\begin{equation}
    \frac{d r_\star}{d r} = \frac{1}{f(r)} \quad \Rightarrow \quad r_\star = \int \frac{1}{f(r)} \, dr.
    \label{tortoise_def}
\end{equation}
The new coordinate, \( r_\star \), is called the tortoise coordinate, which approaches \( r_\star \to -\infty \) at the event horizon, and \( r_\star \to +\infty \) at asymptotic infinity. The differential equation \eqref{sc_radial} in terms of the tortoise coordinate reads,
\begin{equation}
    \frac{\partial^{2} \psi}{\partial t^{2}} - \frac{\partial^{2} \psi}{\partial r_\star^{2}}  + \left(\frac{\ell (\ell+1)}{r^2} + \frac{f'(r)}{r} \right)f(r) \psi  = 0.
    \label{scalar_pert_eqn}
\end{equation}

The scalar field is oscillatory with respect to time with an angular frequency of \( \omega \), which is the quasinormal mode, and is given by \( \psi(t,r_\star) = e^{i\omega t} \psi(r_\star) \footnote{Given the static nature of the background metric, the time dependence of metric perturbations can be decomposed into Fourier modes,  \[ F(t,r)=\int _{-\infty} ^{+\infty} d\omega \tilde F (\omega , r) e^{-i\omega t}. \] Consequently, the time derivatives become \( -i \omega \) in the equations of motion.} \). Substituting this in Eq.~(\ref{scalar_pert_eqn}), we get,
\begin{equation}
    \frac{d^2\psi}{dr_\star^2} + \left[\omega^2 - V_{SC}(r)\right] \psi = 0,
    \label{scalar_sch_eq}
\end{equation}
where \( V_{SC}(r) \) is the effective potential of scalar field perturbation, which is given by,
\begin{equation}
    V_{SC}(r) = \left(\frac{\ell (\ell+1)}{r^2} + \frac{2 M}{r^{3}} + \frac{2 K w}{r^{2(w+1)}}\right) \left(1-\frac{2M}{r}-\frac{K}{r^{2w}}\right).
\end{equation}
This potential for scalar perturbation becomes similar to the potential for scalar perturbation in the background of a Kislev black hole when the equation of state parameter \(w\) and the Kislev parameter \(\epsilon\) are related as $ 2w = 3\epsilon + 1$ and the constants \(K\) (for the anisotropic matter field) and \(c\) (for the Kislev black hole) are equal $K = c$ \cite{Chen:2005qh}.

\subsection{Electromagnetic Perturbation}
Now we focus on the electromagnetic field perturbations, which are described by the vacuum Maxwell's equations in black hole spacetime,
\begin{equation}
    \nabla_\mu F^{\mu\nu} = \frac{1}{\sqrt{-g}} \partial_\mu \left(\sqrt{-g} \,F^{\mu\nu}\right) = 0,
    \label{maxwell_eqn}
\end{equation}
where \( \nabla_\mu \) denotes the covariant derivative. \( F^{\mu\nu} \) is the electromagnetic stress tensor (Faraday tensor) defined in terms of the potential \( A_\mu \) as \( F_{\mu\nu} = \nabla_\mu A_\nu - \nabla_\nu A_\mu \). Since the background spacetime is spherically symmetric, we can use vector spherical harmonics to describe the potential \( A_\mu \), whose components are given by \cite{Cardoso:2001bb, Dey:2018cws},
\begin{equation}
    A_\mu = \sum_{\ell,m} \left[{\begin{pmatrix}0\\0\\ \frac{b^{m}_{\ell}}{\sin(\theta)}\partial_\varphi Y^{m}_{\ell}\\ - b^{m}_{\ell} \sin(\theta)  \partial_\theta Y^{m}_{\ell}\end{pmatrix}} +  \begin{pmatrix} f^{m}_{\ell} Y^{m}_{\ell} \\h^{m}_{\ell}Y^{m}_{\ell} \\k^{m}_{\ell} \partial_\theta Y^{m}_{\ell}\\k^{m}_{\ell} \partial_\varphi Y^{m}_{\ell}\end{pmatrix}\right],
    \label{four_vec_def}
\end{equation}
where \( b^m_\ell, f^m_\ell, h^m_\ell, k^m_\ell \) are functions that depend on the radial and time coordinates \( (t,r) \). Under the angular space inversion transformation \( (\theta , \varphi) \to (\pi - \theta , \pi + \varphi) \), the parity of the first part is \( (-1)^{\ell+1} \), which is the axial or odd term ( magnetic-type parity), and the parity of the second part is \( (-1)^\ell \), which is the polar or even part (electric-type parity). Due to the spherical symmetry of the background spacetime, the axial and polar perturbations can be treated independently and the solution will be independent of $m$. Moreover, both axial and polar equations will lead to identical results \cite{Chandrasekhar:1985kt}. Therefore, we focus on the axial electromagnetic perturbations.

Using the axial part of Eq.~\eqref{four_vec_def}, the independent non-zero components of the Faraday tensor corresponding to axial perturbation are,
\begin{align}
    F_{t \theta } &= \frac{1}{\sin(\theta)} \, \partial _\varphi Y^m_\ell \, \partial _t \psi \nonumber \\
    F_{r \theta } &= \frac{1}{f(r) r^2 \sin(\theta)} \, \partial _\varphi Y^m_\ell \, \partial _r \psi \nonumber \\
    F_{\theta \varphi} &= \ell (\ell+1) \, \sin(\theta) \,  Y^{m}_{\ell} \, \psi \\
    F_{t \varphi} &= - \sin(\theta) \, \partial _\theta Y^m_\ell \, \partial _t \psi \nonumber \\
    F_{\varphi r} &= \sin(\theta) \, \partial _\theta Y^m_\ell \, \partial _r \psi \nonumber
\end{align}
where we denote \( b(t,r) \) by \( \psi(t,r) \). Expanding Eq.~\eqref{maxwell_eqn} by substituting the components of \( F_{\mu\nu} \), and separating the variables, the radial part of the equation in terms of the tortoise coordinate \eqref{tortoise_def} takes the following form,
\begin{equation}
    \frac{\partial^{2}\psi}{\partial t^{2}} - \frac{\partial^{2}\psi}{\partial r_\star^{2}} + \frac{\ell (\ell+1) f(r) }{r^2} \psi = 0.
\end{equation}
Using the Fourier decomposition \( \psi(t,r_\star) = e^{i\omega t}\psi(r_\star) \), we get,
\begin{equation}
    \frac{d^2\psi}{d r_\star^2} + \left[\omega^2 - V_{EM}(r)\right] \psi = 0,
    \label{EM_sch_eq}
\end{equation}
where \( V_{EM} \) is the effective potential of electromagnetic perturbation,
\begin{equation}
    V_{EM}(r) = \frac{\ell (\ell+1)}{r^2} \left(1 - \frac{2M}{r} - \frac{K}{r^{2w}}\right).
\end{equation}

\begin{figure}[t]
\centering
\includegraphics[width=\textwidth]{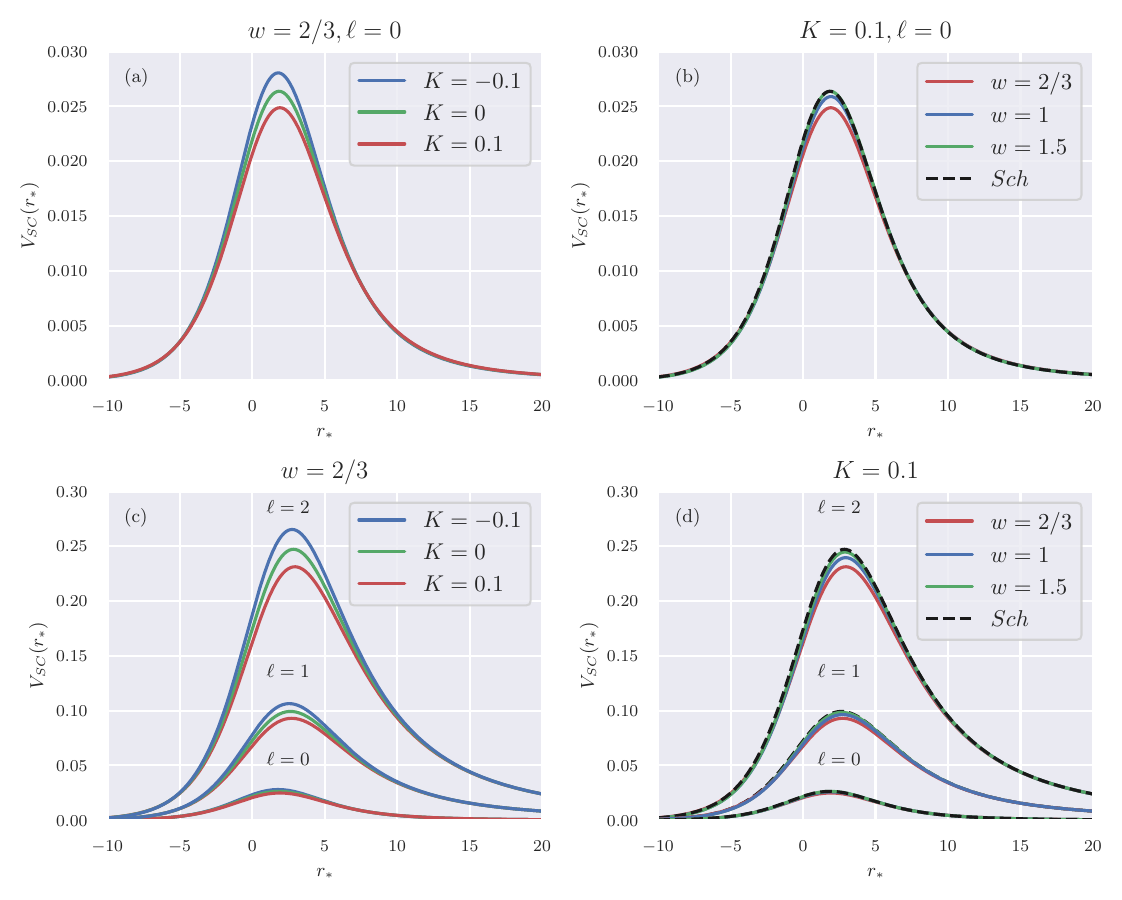}
\caption{The behavior of the effective potential for scalar perturbation. Left panel: variation with \( K \). Right panel: variation with \( w \). In \ref{fig1}(a), the effect of \( K \) is displayed by fixing \( w \). For negative \( K \), the height of the potential increases relative to the Schwarzschild case (\( K=0 \)), whereas for positive \( K \), the height decreases. The strength of this splitting is determined by the \( w \) value, as depicted in \ref{fig1}(b), where it is clear that for smaller \( w \) values, the deviation is more from the Schwarzschild case, i.e., the decrease in the height of the potential is greater for smaller \( w \) values. Similarly, for a fixed negative \( K \) value, the increase in the height of the potential is greater for smaller \( w \) values. In \ref{fig1}(c) and \ref{fig1}(d), it is shown that the observed splitting of the potential exists for all \( \ell \) modes.}
\label{fig1}
\end{figure}

\begin{figure}[t]
\centering
\includegraphics[width=\textwidth]{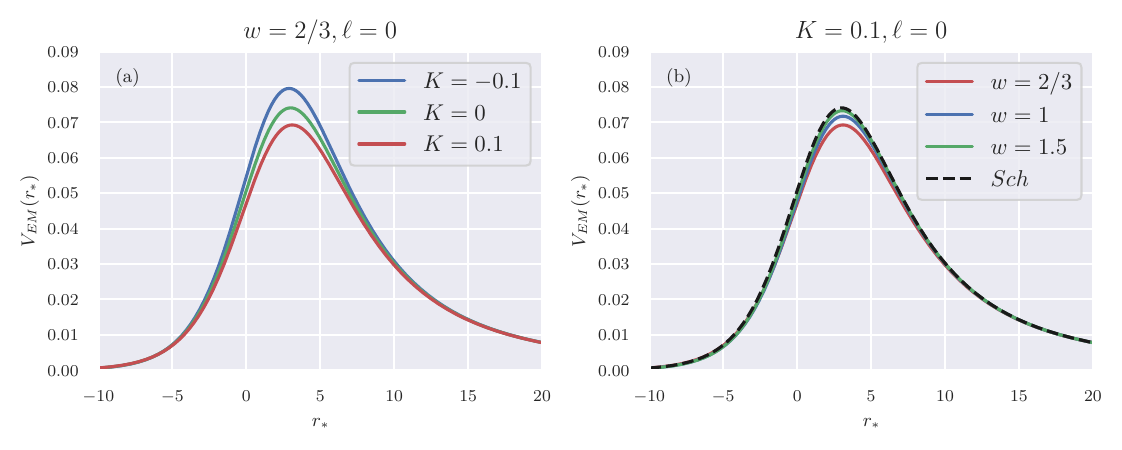}
\caption{The behavior of the effective potential for electromagnetic perturbation. The qualitative behavior of the potential is the same as that of the scalar potential.} 
\label{fig2}
\end{figure}

The effective potential plays an important role in determining the QNM frequencies. We plot the effective potentials for scalar and electromagnetic perturbations in Figs. \ref{fig1} and \ref{fig2}. Both effective potentials approach zero at the horizon and infinity, which is similar to that of the Schwarzschild black hole. In the intermediate region, the behavior of the effective potentials for the perturbations depends on the parameters \( K \) and \( w \). Since both potentials exhibit similar behavior, we chose the scalar case for detailed analysis. 

From Fig. \ref{fig1}(a), we see that for a given value of \( \ell \) and \( w \), there exists a splitting in the effective potential as the \( K \) value is varied, i.e., for positive and negative values of \( K \), the potentials are shifted to either side of the Schwarzschild potential (\( K=0 \)). However, the strength of the splitting depends on the value of \( w \) as shown in Fig. \ref{fig1}(b). We consider only physically admissible values of \( w \), i.e., \( w \geq 1/2 \). It should be noted that for a given value of \( K \) (\( K \neq 0 \)), the limit \( w \to \infty \) corresponds to the Schwarzschild case. 

In Fig. \ref{fig1}(b), we fix \( K \) to a positive value and vary the \( w \) value. The potential is more deviated from that of the Schwarzschild case for smaller values of \( w \) (less anisotropy). As the \( w \) value increases, the potential approaches the Schwarzschild case. The qualitative behavior of the effective potential with varying \( K \) and \( w \) is true for all \( \ell \) modes (see Figs. \ref{fig1}(c) and \ref{fig1}(d)). However, for a given negative value of \( K \), the increase in \( w \) increases the height of the potential.

\section{Quasinormal Modes}\label{sec3}

The QNM frequencies, \( \omega \) in Eqs.~\eqref{scalar_sch_eq} and \eqref{EM_sch_eq}, are computed by requiring appropriate boundary conditions \cite{Konoplya:2011qq}. Near the event horizon, \( r_\star\rightarrow -\infty \), the waves are purely incoming, and at asymptotic infinity, \( r_\star \to \infty \), they are purely outgoing,
\begin{equation}\label{qnmbc}
    \psi(r_\star) \sim e^{i \omega (t \pm r_\star)} \qquad r_\star \to \pm \infty.
\end{equation}
The effective potentials of both the scalar and electromagnetic perturbations exhibit similar behavior with a single peak and vanishing tails at the event horizon and asymptotic infinity. These behaviors of the effective potential are the necessary boundary conditions to use the semi-analytical methods to calculate the QNM frequencies. Here, we will use the WKB method to calculate the QNM frequencies, which was first developed in Ref. \cite{Schutz:1985km}. Using the asymptotic behavior given above and the appropriate boundary conditions (see \cite{Konoplya:2011qq} for details), the first-order solution to the QNM frequencies takes the form,
\begin{equation}
    \omega^2 = V_0 - i \alpha \sqrt{-2V^{(2)}_0}.
\end{equation}
Here, \( \alpha = n + \frac{1}{2} \), where \( n \) is the overtone number, which takes the values \( n=0,1,2,3,\ldots \), and \( V_0 = V(r_{\star_{0}}) \), where \( r_{\star_{0}} \) is the critical point (the maximum of the effective potential in tortoise coordinates defined as \( \frac{d V}{d r_\star}\big|_{r_\star=r_{\star_{0}}}=0 \)), and \( V^{(m)}_0 = \frac{d^m V}{d r_\star^m}\big|_{r_{\star_{0}}} \) is the value of the \( m \)-th derivative of \(V\) at \( r_{\star_{0}} \). The analytical expression for \( r_{\star_{0}} \) is not always feasible, making the WKB method semi-analytical. The precision of the WKB method can be improved by calculating higher-order terms. 

The third-order WKB approximation formula given in \cite{Iyer:1986np} is shown below,
\begin{equation}
\begin{split}
     \omega^2 = & \left[V_0 + \sqrt{ -2V^{(2)}_0}\Lambda_2\right] - i \alpha \sqrt{ -2V^{(2)}_0}(1+\Lambda_3),\\
     \Lambda_2 = & \frac{1}{\sqrt{ -2V^{(2)}_0}}\left[ \frac{1}{8} \left(\frac{V^{(4)}_0}{V^{(2)}_0}\right)\left(\frac{1}{4}+\alpha^2\right) - \frac{1}{288}\left(\frac{V^{(3)}_0}{V^{(2)}_0}\right)^2(7+60\alpha^2) \right], \\
     \Lambda_3 = & \frac{1}{\sqrt{ -2V^{(2)}_0}}\left[ \frac{5}{6912} \left(\frac{V^{(3)}_0}{V^{(2)}_0}\right)^4 (77+188\alpha^2)-\frac{1}{384}\left(\frac{\left(V^{(3)}_0\right)^2V^{(4)}_0}{\left(V^{(2)}_0\right)^3}\right)(51+100\alpha^2) \right.\\
   & \left. +\frac{1}{2304} \left(\frac{V^{(4)}_0}{V^{(2)}_0}\right)^2 (67+68\alpha^2)+\frac{1}{288}\left(\frac{V^{(3)}_0 V^{(5)}_0}{\left(V^{(2)}_0\right)^2}\right)(19+28\alpha^2) -\frac{1}{288} \left(\frac{V^{(6)}_0}{V^{(2)}_0}\right) (5+4\alpha^2)\right].
\end{split}
\end{equation}
Similarly, the 6th order WKB formula is \cite{Konoplya:2003ii},
\begin{equation}
    \frac{i(\omega^2 - V_0)}{\sqrt{-2V^{(2)}_0}} = \alpha + \sum_{i=2}^{6} \Lambda_i.
\end{equation}
The correction terms \( \Lambda_2 \) and \( \Lambda_3 \) are as shown before, and \( \Lambda_4 \), \( \Lambda_5 \), and \( \Lambda_6 \) are given in \cite{Konoplya:2003ii}. Recently, the method has been extended to even higher orders, up to the 13th order, in Ref. \cite{Matyjasek:2017psv}. However, it should be noted that the higher order WKB approximation does not always converge and results in more accuracy, and the optimal order WKB approximation depends on the spacetime potential considered \cite{Konoplya:2019hlu}. The error in WKB order can be estimated by using,
\begin{equation}
    \Delta_i = \frac{|\omega_{i+1} - \omega_{i-1}|}{2},
    \label{wkberroreq}
\end{equation}
where \( \omega_i \) represents the QNM value obtained from the \( i^{\text{th}} \) order WKB approximation. We chose QNM frequencies from the optimal WKB order having the least error.

\begin{table}[ht!]
\centering
\begin{tabular}{@{}m{4.5em} m{14em} m{9em} m{7.4em} m{7em}@{}}
\( K \) & \(\omega _R + i \omega _I \) & \(\Delta _i\) & \(\delta_{\omega_R}\) & \(\delta_{\omega_I}\) \\
\hline
\multicolumn{5}{c}{\textbf{\( l=0 \)}} \\
\hline
$0.2$   & 0.11(0542) - 0.10(4256)\,i [7] & 0.003236 & 2.107817 & 0.729688 \\
$0.1$   & 0.11(1850) - 0.10(3934)\,i [7] & 0.003016 & 0.949399 & 0.418916 \\
$0.01$  & 0.11(2829) - 0.10(3550)\,i [7] & 0.002944 & 0.082948 & 0.047991 \\
$0.001$ & 0.11(2913) - 0.10(3506)\,i [7] & 0.002942 & 0.008167 & 0.004863 \\
$0$      & 0.11(2922) - 0.10(3501)\,i [7] & 0.002942 & 0        & 0        \\
$-0.001$  & 0.11(2932) - 0.10(3496)\,i [7] & 0.002943 & 0.008139 & 0.004877 \\
$-0.01$   & 0.11(3013) - 0.10(3450)\,i [7] & 0.002942 & 0.080125 & 0.049399 \\
$-0.1$    & 0.11(3690) - 0.10(2929)\,i [7] & 0.002986 & 0.679531 & 0.551994 \\
$-0.2$    & 0.11(4259) - 0.10(2292)\,i [7] & 0.003004 & 1.183221 & 1.168214 \\
\hline
\multicolumn{5}{c}{\textbf{\( l=1 \)}} \\
\hline
$0.2$   & 0.2895(73) - 0.0983(49)\,i [8] & 0.000029 & 1.147384 & 0.724132 \\
$0.1$   & 0.2912(24) - 0.0980(10)\,i [8] & 0.000021 & 0.583562 & 0.377308 \\
$0.01$  & 0.2927(60) - 0.0976(80)\,i [9] & 0.000020 & 0.059280 & 0.039157 \\
$0.001$ & 0.2929(16) - 0.0976(46)\,i [9] & 0.000020 & 0.005936 & 0.003945 \\
$0$      & 0.2929(34) - 0.0976(42)\,i [9] & 0.000020 & 0        & 0        \\
$-0.001$  & 0.2929(51) - 0.0976(38)\,i [9] & 0.000020 & 0.005938 & 0.003952 \\
$-0.01$   & 0.2931(08) - 0.0976(03)\,i [9] & 0.000019 & 0.059454 & 0.039813 \\
$-0.1$    & 0.2946(99) - 0.0972(23)\,i [9] & 0.000023 & 0.602565 & 0.428862 \\
$-0.2$    & 0.2965(22) - 0.0967(43)\,i [8] & 0.000032 & 1.224997 & 0.921042 \\
\hline
\multicolumn{5}{c}{\textbf{\( l=2 \)}} \\
\hline
$0.2$   & 0.478356 - 0.097413\,i [10]     & 2.41258$\times 10^{-7}$ & 1.093430 & 0.676781 \\
$0.1$   & 0.480949 - 0.097107\,i [11]     & 3.49189$\times 10^{-7}$ & 0.557133 & 0.359756 \\
$0.01$  & 0.483370 - 0.096795\,i [12]     & 2.10014$\times 10^{-7}$ & 0.056685 & 0.038085 \\
$0.001$ & 0.483616 - 0.096762\,i [12]     & 2.00961$\times 10^{-7}$ & 0.005679 & 0.003831 \\
$0$      & 0.483644 - 0.096759\,i [12]     & 2.00043$\times 10^{-7}$ & 0        & 0        \\
$-0.001$  & 0.483671 - 0.096755\,i [12]     & 1.99142$\times 10^{-7}$ & 0.005681 & 0.003836 \\
$-0.01$   & 0.483919 - 0.096721\,i [12]     & 1.91803$\times 10^{-7}$ & 0.056910 & 0.038582 \\
$-0.1$    & 0.486447 - 0.096362\,i [12]     & 1.93677$\times 10^{-7}$ & 0.579515 & 0.409453 \\
$-0.2$    & 0.489367 - 0.095910\,i [11]     & 2.63226$\times 10^{-7}$ & 1.183352 & 0.876847 \\
\hline
\end{tabular}
\caption{Scalar QNM values obtained from the higher-order WKB method by varying the \( K \) value for different \( l \) values. The significant digit errors are shown in brackets. The optimal WKB order used for each calculation is indicated within the QNM values in square brackets. The parameters fixed are \( M=1 \), \( w=3/2 \), \( n=0 \).}
\label{tabscqnm}
\end{table}

\begin{table}[ht!]
\centering
\begin{tabular}{@{}m{4.5em} m{14em} m{9em} m{7.4em} m{7em}@{}}
\( K \) & \(\omega _R + i \omega _I \) & \(\Delta _i\) & \(\delta_{\omega_R}\) & \(\delta_{\omega_I}\) \\
\hline
\multicolumn{5}{c}{\textbf{\( l=1 \)}} \\
\hline
$0.2$ & 0.2446(35) - 0.0930(01)\,i [6] & 0.000040 & 1.451310 & 0.568229 \\
$0.1$ & 0.2463(74) - 0.0928(40)\,i [6] & 0.000088 & 0.750926 & 0.394419 \\
$0.01$ & 0.2480(83) - 0.0924(65)\,i [12] & 0.000042 & 0.062609 & 0.011413 \\
$0.001$ & 0.2482(22) - 0.0924(75)\,i [12] & 0.000032 & 0.006280 & 0.001100 \\
$0$ & 0.2482(38) - 0.0924(76)\,i [12] & 0.000035 & 0 & 0 \\
$-0.001$ & 0.2482(54) - 0.0924(77)\,i [12] & 0.000038 & 0.006285 & 0.001091 \\
$-0.01$ & 0.2484(23) - 0.0924(75)\,i [11] & 0.000030 & 0.074559 & 0.001001 \\
$-0.1$ & 0.2502(42) - 0.0921(56)\,i [9] & 0.000058 & 0.807194 & 0.345679 \\
$-0.2$ & 0.2522(14) - 0.0917(35)\,i [9] & 0.000085 & 1.601690 & 0.801271 \\
\hline
\multicolumn{5}{c}{\textbf{\( l=2 \)}} \\
\hline
$0.2$ & 0.45219(6) - 0.09561(0)\,i [6] & 1.582946$\times 10^{-6}$ & 1.180020 & 0.636960 \\
$0.1$ & 0.454843 - 0.095326\,i [12] & 9.000555$\times 10^{-7}$ & 0.601523 & 0.338649 \\
$0.01$ & 0.457315 - 0.095039\,i [12] & 9.682657$\times 10^{-8}$ & 0.0612684 & 0.036037 \\
$0.001$ & 0.457567 - 0.095008\,i [12] & 5.300744$\times 10^{-8}$ & 0.006138 & 0.003626 \\
$0$ & 0.457595 - 0.095004\,i [12] & 5.349791$\times 10^{-8}$ & 0 & 0 \\
$-0.001$ & 0.457624 - 0.095001\,i [12] & 5.540361$\times 10^{-8}$ & 0.006140 & 0.003631 \\
$-0.01$ & 0.457877 - 0.094970\,i [12] & 1.089834$\times 10^{-7}$ & 0.061524 & 0.036547 \\
$-0.1$ & 0.460466 - 0.094633\,i [10] & 4.472695$\times 10^{-7}$ & 0.627339 & 0.390971 \\
$-0.2$ & 0.463463 - 0.094205\,i [9] & 6.848180$\times 10^{-7}$ & 1.282340 & 0.841890 \\
\hline
\multicolumn{5}{c}{\textbf{\( l=3 \)}} \\
\hline
$0.2$ & 0.649529 - 0.096236\,i [12] & $3.627429\times 10^{-8}$ & 1.121858 & 0.648604 \\
$0.1$ & 0.653141 - 0.095947\,i [12] & $1.782784\times 10^{-8}$ & 0.571957 & 0.345698 \\
$0.01$ & 0.656516 - 0.095651\,i [12] & $2.118348\times 10^{-9}$ & 0.058242 & 0.036678 \\
$0.001$ & 0.656860 - 0.095620\,i [12] & $1.000391\times 10^{-9}$ & 0.005835 & 0.003689\\
$0$ & 0.656899 - 0.095616\,i [12] & $9.529588\times 10^{-10}$ & 0 & 0 \\
$-0.001$ & 0.656937 - 0.095613\,i [12] & $9.348805\times 10^{-10}$ & 0.005837 & 0.003695\\
$-0.01$ & 0.657283 - 0.095581\,i [12] & $1.783486\times 10^{-9}$ & 0.058481 & 0.037172 \\
$-0.1$ & 0.660814 - 0.095238\,i [10] & $1.408763\times 10^{-8}$ & 0.595962 & 0.395305 \\
$-0.2$ & 0.664901 - 0.094805\,i [12] & $3.123030\times 10^{-8}$ & 1.218143 & 0.848420 \\
\hline
\end{tabular}
\caption{Electromagnetic QNM values obtained from higher-order WKB method by varying the \( K \) value for different \( l \) values. The significant digit error is shown in brackets. In each case, we have chosen the optimal WKB order, which is indicated in square brackets. The parameters fixed are \( M=1 \), \( w=3/2 \), \( n=0 \).}
\label{tabemqnm}
\end{table}

We have calculated QNM frequencies for both scalar and electromagnetic perturbations for different values of \( w \) and \( K \) in Tables \ref{tabscqnm} and \ref{tabemqnm}. The details of error analysis are given in Appendix \ref{errapp}. We find that the error in the WKB approximation is always negligible compared to the effect of the anisotropic matter field on both the imaginary and real parts of the QNM frequencies, as shown in Fig. \ref{fig3}. In the QNM frequencies, we observe that for a given value of \( w \), the change in \( K \) induces a splitting around Schwarzschild QNMs. This is in direct correlation with the splitting observed in the effective potential. 

It is interesting to note that this behavior is analogous to the Zeeman-like splitting observed in other black hole QNM studies. For example, a Zeeman-like splitting is observed in slowly rotating Kerr black holes \cite{Leaver:1985ax, Konoplya:2011qq} and non-commutative black hole perturbations \cite{Ciric:2017rnf, DimitrijevicCiric:2019hqq, Herceg:2023pmc}. However, the observed splitting in the presence of an anisotropic matter field is not exactly a Zeeman-like splitting as there is no azimuthal quantum number \( m \) involved. Although there is an analogy between the RN black hole and a black hole surrounded by an anisotropic matter field, the splitting we observe in the latter makes it distinct from the former (see Ref. \cite{Kokkotas:1988fm} for the charged case). The Schwarzschild case corresponds to either \( K=0 \) or \( w \to \infty \). 

\begin{figure}[t]
\centering
\includegraphics[width=\textwidth]{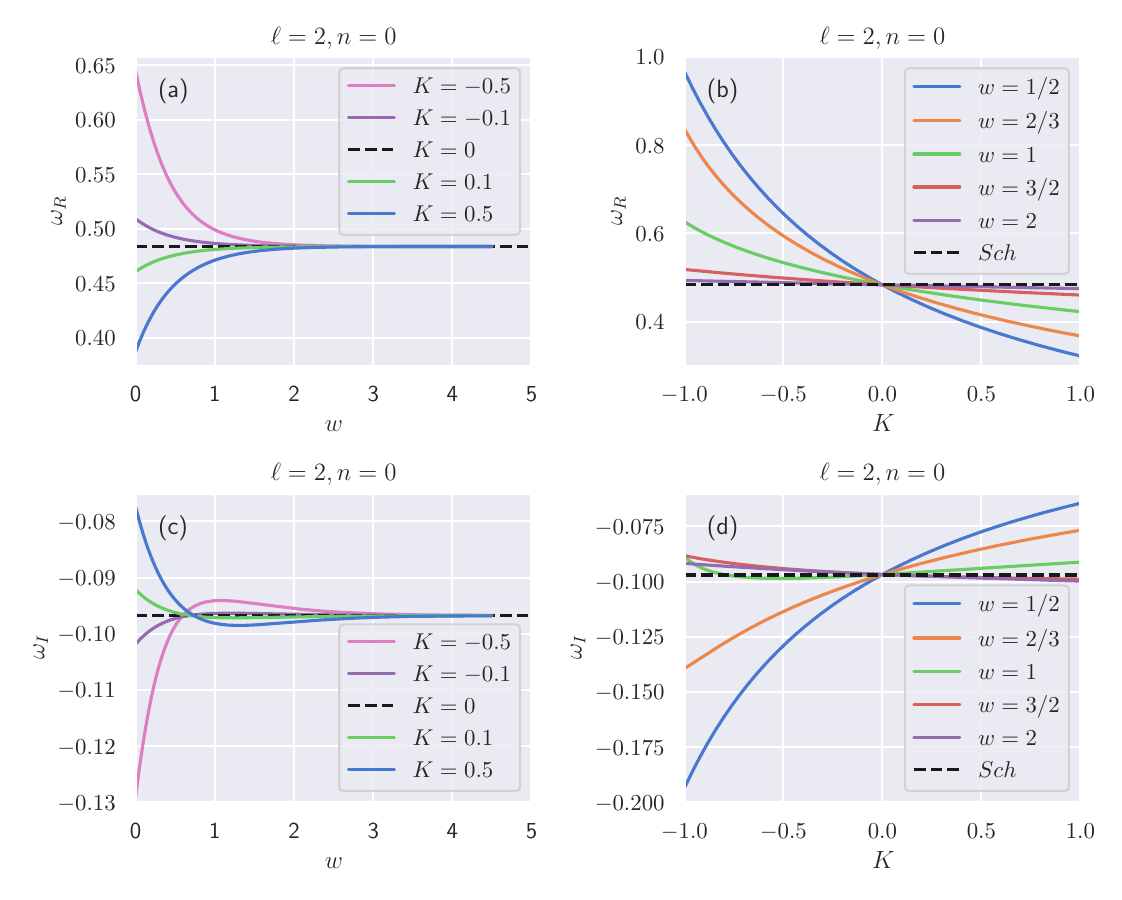}
\caption{Left panel: Effect of \( w \) on QNM frequencies. Right panel: Effect of \( K \) on QNM frequencies. The angular mode \( l = 2 \) is chosen as it is the lowest mode where the error in WKB method is significantly smaller than the corrections introduced by the anisotropic matter field.}
\label{fig4}
\end{figure}

We present a detailed analysis of the effect of both parameters in Fig. \ref{fig4}. The horizontal line (black dotted line) corresponds to the Schwarzschild QNM values. From Fig. \ref{fig4}(a), it is clear that the curves corresponding to negative and positive \( K \) values lie above and below the Schwarzschild line, and they all converge to the Schwarzschild line as \( w \) increases. The strength of deviation depends on the \( w \) value; the smaller the \( w \), the higher the strength of the splitting \ref{fig4}(b). The splitting of QNMs is flipped for the imaginary QNM compared to the real case (Figs. \ref{fig4}(c) and \ref{fig4}(d)). For negative values of \( K \) (with $w$ value fixed to $3/2$), the real and imaginary parts of the QNM frequencies seem to follow the same pattern as for the charged black hole (see Ref. \cite{Kokkotas:1988fm, Leaver:1990zz}). However, it is important to note that this argument cannot be generalized for arbitrary $w$ values due to the crossover observed in Figs.~\ref{fig4}(c) and \ref{fig4}(d). The splitting of QNM values in the presence of an anisotropic matter field is also evident from the numerical values provided in Tables \ref{tabscqnm} and \ref{tabemqnm}. \\

Since the values of \( K \) and \( w \) characterize different fluid configurations as discussed in Section \ref{sec1}, it can be argued that the deviated QNM frequencies are, in fact, a measure of the fluid properties around a Schwarzschild black hole. This deviation from the Schwarzschild black hole can be calculated using the relative effect defined as \cite{Churilova:2020bql},
\begin{equation}
    \delta_{\omega_R} = \frac{\abs{\omega_{R_a}-\omega_{R_{s}}}}{\omega_{R_{s}}}\times 100 \%
\end{equation}
\begin{equation}
    \delta_{\omega_I} = \frac{\abs{\omega_{I_a}-\omega_{I_{s}}}}{\omega_{I_{s}}}\times 100 \%
\end{equation}
where $\omega_{R_a}$ and $\omega_{I_a}$ are the value of real and imaginary parts of QNMs in the presence of the anisotropic matter field. $\omega_{R_{s}}$ and $\omega_{I_{s}}$ are that of schwarzchild limit. The values $\delta_{\omega_R}$ and $\delta_{\omega_I}$ are listed in Tables \ref{tabscqnm} and \ref{tabemqnm}, from which it is evident that the splitting of QNM values around the Schwarzschild values is not symmetric.

\section{Critical Orbits around Black hole and their Connection to QNMs}\label{sec4}

In this section, we study the shadow radius and Lyapunov exponent and their relation with the quasinormal modes we analyzed previously. The study of geodesic motion is important in determining the observable aspects of black hole spacetimes. Particularly, the unstable null geodesics are of importance as they are closely related to the optical properties of black holes and are also associated with the quasinormal modes, as the angular velocity determines the real part of QNM and the instability timescale of the orbit is related to the imaginary part of QNM. In Ref. \cite{Cardoso:2008bp}, it was shown that in the eikonal limit, for any static, spherically symmetric, asymptotically flat spacetime, the angular velocity \( \Omega _c \) at the unstable photon orbit and the Lyapunov exponent \( \lambda \), which determines the instability timescale of the orbit, are related to the analytic WKB approximations for QNMs as
\begin{equation}\label{lyapunov_eq}
    \omega _{QNM} = \Omega _c \ell - i \left( n + \frac{1}{2} \right) |\lambda|
\end{equation}
where \( n \) is the overtone number and \( \ell \) is the angular momentum of the perturbation. The eikonal (geometric optics) limit is valid for a wide class of potentials associated with the massless perturbations, including the scalar and electromagnetic perturbations, which have the same behavior in the eikonal limit. In the eikonal limit, the real part of QNM is related to the shadow radius and the imaginary part to the Lyapunov exponent.

To obtain the connection between the unstable null geodesic and quasinormal mode, we consider the Lagrangian of particles,
\begin{equation}
    \mathcal{L} = \frac{1}{2} g_{\mu\nu} \dot{x}^\mu \dot{x}^\nu,
\end{equation}
where the dot denotes the derivative with respect to the affine parameter \( s \). The conjugate momenta \( p_\mu \) are given by \( p_\mu = \partial \mathcal{L} / \partial \dot{x}^\mu \). Due to the spacetime symmetry, the conjugate momenta \( p_t \) and \( p_\varphi \) are conserved quantities which define the total energy \( E \) and the angular momentum \( L \). Thus we have,
\begin{align}
    p_t &= - f(r)\dot{t} = -E, \label{conjugate_p_t} \\
    p_\varphi &= r^2\sin^2(\theta) \dot{\varphi} = L. 
    \label{conjugate_p_phi}
\end{align}
Using the above equations, the Hamiltonian \(\mathcal{H}=p_\mu \dot{x}^\mu - \mathcal{L}\) reads,
\begin{equation}
    \mathcal{H} = \frac{1}{2}\left[-\frac{E^2}{f(r)} + \frac{\dot{r}^2}{f(r)} + r^2\dot{\theta}^2 + \frac{L^2}{r^2\sin^2(\theta)}\right].
\end{equation}
Since the spacetime is spherically symmetric, we restrict the analysis to the equatorial plane (\(\theta = \pi/2\)) without loss of generality. For null geodesics, \(ds^2 = 0\), which implies \(\mathcal{H} = 0\) and we get \(\dot{r}^2 + V_\text{eff}(r) = 0\), where \(V_\text{eff}(r)\) is the effective potential,
\begin{equation}
    V_\text{eff}(r) = E^2 - \frac{L^2 f(r)}{r^2}.
    \label{geodesic_veff}
\end{equation}

\subsection{Shadow Radius and QNM}

\begin{figure}[t]
\centering
\includegraphics[width=\textwidth]{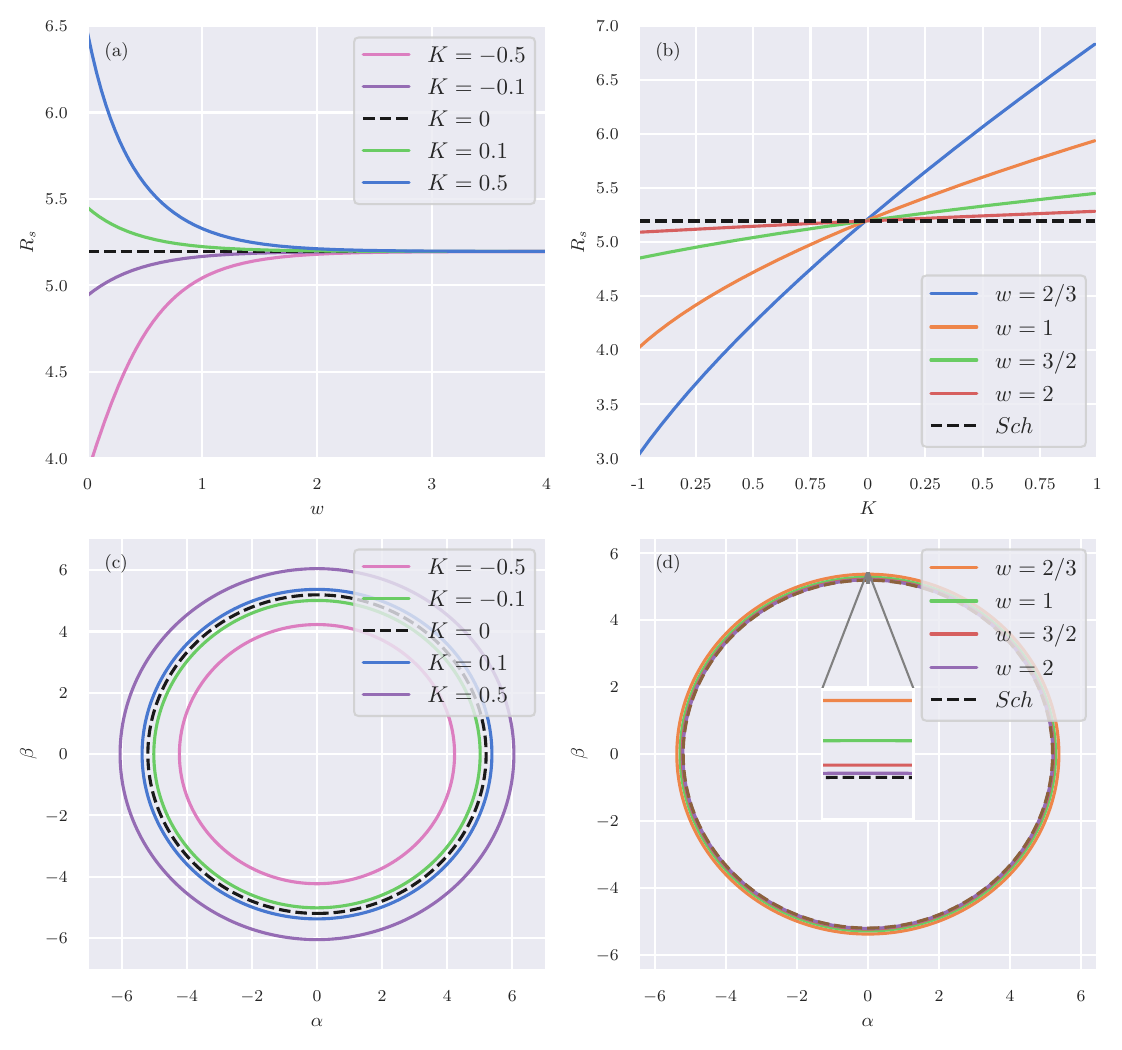}
\caption{(a) Effect of \( w \) on the shadow radius. (b) Effect of \( K \) on the shadow radius. (c) Circles representing the shadow radius in celestial coordinates for various \( K \) values, with \( w \) fixed at \( w = 2/3 \). (d) Circles representing the shadow radius in celestial coordinates for various \( w \) values, with \( K \) fixed at \( K = 0.1 \).}
\label{fig5}
\end{figure}

The photon sphere, the innermost circular orbit, comprises unstable photon orbits that are in close vicinity to the black hole event horizon, defining the boundary of the shadow cast by a compact object. For the innermost circular orbit, \( V_\text{eff}(r) = V_{\text{eff}}'(r) = 0 \) (where prime \( ' \) denotes the derivative with respect to the radial coordinate \( r \)), which gives,
\begin{equation}
    rf'(r) - 2f(r) = 0
\end{equation}
and
\begin{equation} \label{impeqn}
    \frac{L}{E} = \frac{r}{\sqrt{f(r)}} \equiv \xi
\end{equation}
where the quantity \( \xi \) is called the impact parameter. Photons originating from a distant source with an impact parameter \( \xi \) greater than the critical impact parameter, \( \xi_c \), remain outside the photon sphere and reach the observer. Conversely, photons with impact parameters smaller than \( \xi_c \) are trapped within the photon sphere and do not reach the observer, creating dark regions in the observer's sky. The aggregation of these dark regions forms the shadow \cite{Synge:1966okc, Luminet:1979nyg}. The critical impact parameter (\( \xi_c \)) at \( r = r_p \) (radius of the photon sphere) is the shadow radius of the black hole \( R_s \). Hence we get,
\begin{equation} \label{shadow_r}
    R_s = \frac{r_p}{\sqrt{f(r_p)}} \equiv \sqrt{\alpha^2 + \beta^2}
\end{equation}
where \( (\alpha, \beta) \) are the celestial coordinates. Using this equation, we studied the behavior of the shadow radius with \( K \) and \( w \) in Fig. \ref{fig5}. The shadow radius increases from the Schwarzschild value for positive values of \( K \), whereas it decreases for negative values. Physically, this implies that an anisotropic matter field surrounding a Schwarzschild black hole influences the shadow radius. We find that the qualitative nature of the plots is similar to that of the real part of QNMs in Fig. \ref{fig4}. Particularly, the splitting behavior has a striking similarity to the plots of the real part of QNMs in Fig. \ref{fig4}.

On the other hand, there exists a relationship between gravitational lensing in the strong-deflection limit and the frequencies of the quasinormal modes, which is given as \cite{Stefanov:2010xz},
\begin{align} \label{lens_eqn}
    \Omega _c = \frac{c}{\theta D_{OL}}, \qquad \lambda = \frac{c \ln \tilde{r}}{2\pi \theta D_{OL}},
\end{align}
where \( c \) is the speed of light, \( \theta = R_s/D \) is the angular position of the image, \( D_{OL} \) is the distance from the observer to the lens, and \( \tilde{r} \) is the flux ratio. Using \eqref{lens_eqn}, \eqref{shadow_r}, and \eqref{lyapunov_eq}, the relationship between the eikonal QNM and shadow radius is established in Ref \cite{Jusufi:2019ltj},
\begin{equation} \label{omegaR_eq}
    \omega _R = \lim_{\ell \gg 1} \frac{\ell}{R_s}. 
\end{equation}
We have studied the relationship between the shadow radius and the real part of QNM in Fig. \ref{fig7}(a), from which it is evident that in the eikonal limit, they exactly match as given in \eqref{omegaR_eq}. The relationship exists for all values of \( K \) and \( w \).

\subsection{Lyapunov Exponent and QNM}

\begin{figure}[t!]
\centering
\includegraphics[width=\textwidth]{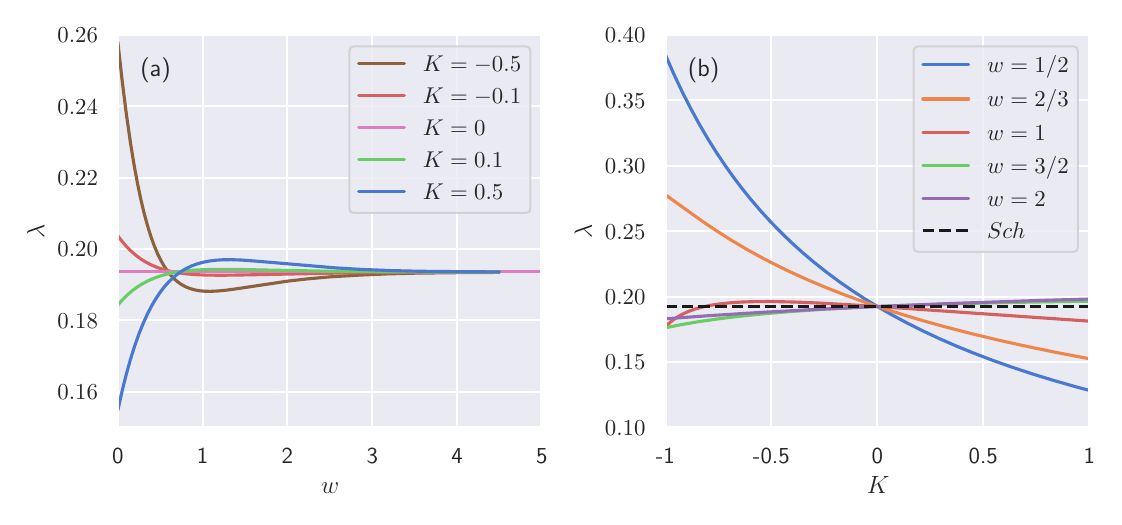}
\caption{Effect of \( w \) (left) and \( K \) (right) on Lyapunov Exponent.} 
\label{Lyapunov_plot}
\end{figure}

The Lyapunov exponent \( \lambda \) quantifies the average rate at which nearby trajectories in phase space either converge or diverge. If \( \lambda > 0 \), it signifies divergence between nearby trajectories, indicating a high sensitivity to initial conditions. For a spherically symmetric spacetime, the null geodesic stability analysis using the Lyapunov exponent results in \cite{Cardoso:2008bp, Kumara:2024obd}
\begin{equation}
    \lambda = \sqrt{\frac{V^{\prime\prime}_\text{eff}(r_p)}{2\dot{t}^2}} 
    \label{def_of_lyapunov}
\end{equation}
where \( V_\text{eff}(r) \) is given in \eqref{geodesic_veff}. For an unstable circular geodesic, we have \( V_\text{eff}(r) = 0 \), \( V^\prime_\text{eff}(r) = 0 \), and \( V^{\prime\prime}_\text{eff}(r) > 0 \). From \eqref{geodesic_veff},
\begin{equation}
    V^{\prime\prime}_\text{eff}(r_p) = \frac{L^2}{r_p^4}\big[2 f(r_p) - r_p^2 f''(r_p)\big] .
    \label{V_eff_rp}
\end{equation}
Substituting \eqref{V_eff_rp} into \eqref{def_of_lyapunov}, and using \eqref{conjugate_p_t} and \eqref{impeqn}, we obtain,
\begin{equation}
    \lambda = \sqrt{\frac{f(r_p)\big[2 f(r_p) - r_p^2 f''(r_p)\big]}{2r_p^2}} .
\end{equation}
We studied the behavior of the Lyapunov exponent using this equation in Fig. \ref{Lyapunov_plot} for different values of \( K \) and \( w \). We find that the behavior is analogous to the imaginary part of QNM as shown in Fig. \ref{fig4}.

\begin{figure}[t!]
\centering
\includegraphics[width=\textwidth]{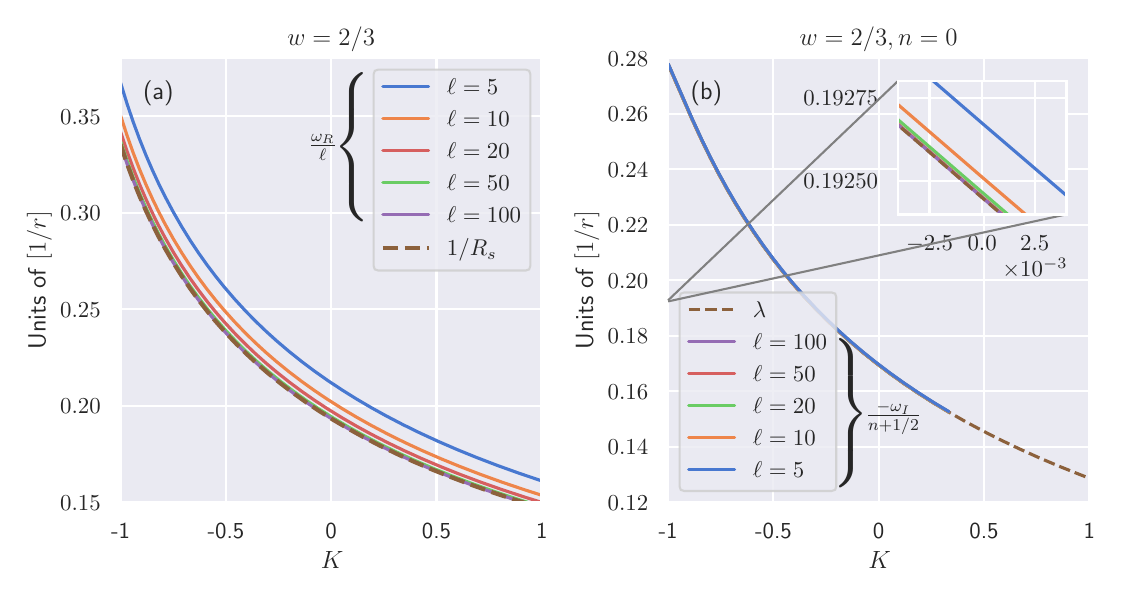}
\caption{Left: Relation between the shadow radius and real part of QNM in the eikonal limit. Right: Relation between the Lyapunov Exponent and imaginary part of QNM in the eikonal limit. } 
\label{fig7}
\end{figure}

On the other hand, there exists a relationship between the Lyapunov exponent and the imaginary part of QNMs in the eikonal limit as shown in \eqref{lyapunov_eq},
\begin{equation}
    \lambda = \lim_{\ell \gg 1}\left( -\frac{\omega_I}{\left(n + \frac{1}{2}\right)} \right) .
    \label{eikonal_lyapunov}
\end{equation}
We studied the connection between the Lyapunov exponent and the imaginary part of QNMs in Fig. \ref{fig7}(b), and we find that in the eikonal limit they match according to this equation. The relationship exists for all values of \( K \) and \( w \).

\section{Scattering and Grey-Body Factors}\label{sec5}

In this section, we analyze reflection coefficients \( R(\omega) \) and transmission coefficients \( T(\omega) \) for various values of black hole parameters. Radiation originating near a black hole event horizon is emitted into the surrounding space, necessitating traversal through a non-trivial, curved spacetime geometry before detection by an observer, such as one situated at asymptotic infinity in an asymptotically flat spacetime. Consequently, the surrounding spacetime acts as a potential barrier for the radiation, resulting in a deviation from the original radiation spectrum observed by an asymptotic observer. This deviation, quantified by the relative factor between the asymptotic radiation spectrum and the emitted radiation spectrum, is commonly referred to as the grey-body factor. Understanding this concept is straightforward by examining the behavior of the effective potential. For both scalar and electromagnetic perturbations, a finite potential barrier (Figs. \ref{fig1} and \ref{fig2}) exists between the event horizon and asymptotic infinity. As a result, any wave traversing black hole spacetime encounters these finite positive potential barriers as obstacles, leading to partial reflection and transmission of the wave. As demonstrated previously, the radial perturbation equations can be reduced to a Schr\"{o}dinger-like equation, which characterizes the scattering of waves in black hole spacetime. This equation yields asymptotic solutions given by,
\begin{align}
    \psi (r_\star) & =  T(\omega) e^{-i\omega r_\star}, &  r_\star \to -\infty \\
    \psi (r_\star) & =  e^{-i\omega r_\star} + R(\omega) e^{i\omega r_\star}, &  r_\star \to +\infty    
\end{align}
Note that these conditions are different from the one we used in the calculation of QNMs \eqref{qnmbc}. Next, we proceed to compute the square of the amplitude of the wave function, which is partially transmitted and partially reflected by the potential barrier. The conservation of probability dictates that,
\begin{equation}
    |R(\omega)|^2 + |T(\omega)|^2 = 1 .
    \label{conveq}
\end{equation}
We employ the WKB approximation to calculate the transmission and reflection coefficients, which provide reasonable accuracy \cite{Konoplya:2020cbv, Konoplya:2020jgt, Konoplya:2023ppx}. Specifically, the reflection amplitude is expressed as \cite{Iyer:1986np},
\begin{equation}
    R(\omega) = \frac{1}{\sqrt{1 + e^{-2\pi i \alpha}}}.
    \label{reflection_coeff}
\end{equation}
Considering higher-order WKB approximations, \(\alpha\) in the above equation is given by \cite{Konoplya:2023ppx},
\begin{equation}
    \alpha = \frac{i(\omega^2 - V_0)}{\sqrt{-2V^{(2)}_0}} - \sum_{i=2}^{6} \Lambda_i (\alpha)
    \label{phase_factor}
\end{equation}
where \( V_0 \) is the peak value of the effective potential, \( V^{(2)}_0 \) is the value of the second derivative with respect to the tortoise coordinate, and \( \Lambda_i(\alpha) \) are the higher-order WKB corrections. The grey-body factor is defined as \( \gamma_{\ell} = |T(\omega)|^2 \). Substituting \eqref{reflection_coeff} and \eqref{phase_factor} into \eqref{conveq}, we obtain,
\begin{equation}
    \gamma_{\ell} = |T(\omega)|^2 = 1 - \left| \frac{1}{\sqrt{1 + e^{-2\pi i \alpha}}} \right|^2.
\end{equation}

Depending on the frequency and height of the potential barrier, various scenarios may arise. When \( \omega^2 \gg V_0 \), indicating that the frequency of the wave is significantly greater than the height of the barrier, reflection of the wave by the barrier does not occur. Consequently, one would anticipate the reflection coefficient to approach zero, as the wave frequency permits it to cross the barrier. Hence, under these conditions, \( |T(\omega)|^2 \) is expected to be close to 1. Conversely, when \( \omega^2 \ll V_0 \), signifying that the frequency is much smaller than the barrier height, the wave will be reflected back by the barrier. Additionally, depending on the values of \( \omega \) and \( V_0 \), some portion of the wave may tunnel through the barrier. In this scenario, we should observe precisely the opposite behavior of \( R(\omega) \) and \( T(\omega) \) compared to the previous case.

\begin{figure}[t]
\centering
\includegraphics[width=\textwidth]{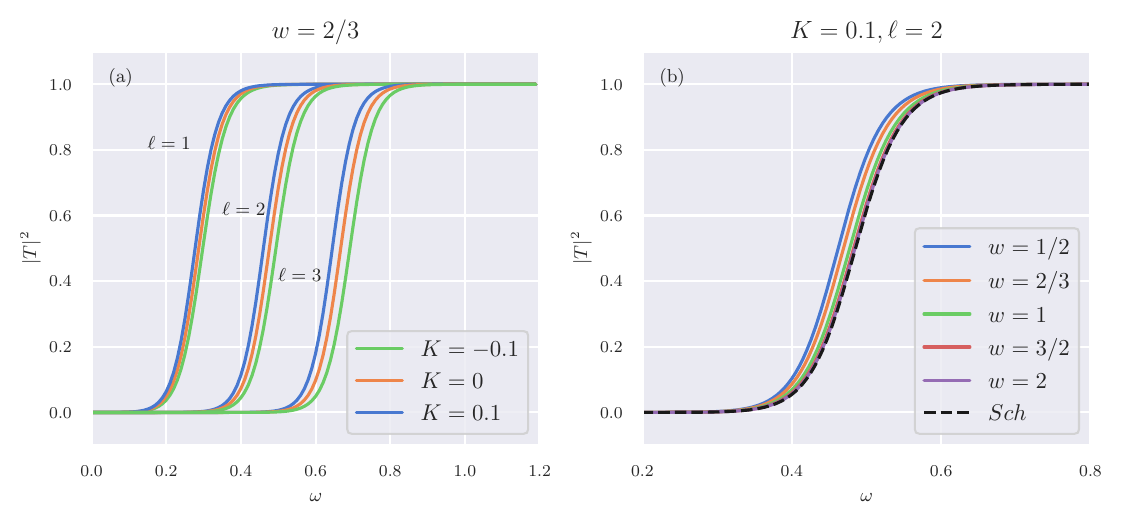}
\caption{Effect of \( K \) (left) and \( w \) (right) on grey-body factor for scalar perturbation. The effect of \( \ell \) is also shown on the left.} 
\label{greybody}
\end{figure}

From Fig. \ref{greybody}, the splitting behavior of the grey-body factor, similar to QNM frequencies, is evident. Here we have considered only scalar perturbation. The electromagnetic perturbation also exhibits similar behavior. It can be seen that the larger the \( K \), the larger the grey-body factor, meaning that a smaller portion of particles is reflected by the effective potential. This can be easily understood from the behavior of the effective potential, which becomes lower (i.e., easier to penetrate) for larger \( K \). However, the strength of the enhancement or suppression of the grey-body factor depends on \( w \). This phenomenon exists for different \( \ell \) modes. The dependence of the grey-body factor on \( \ell \) is the same as that of the Schwarzschild case.

Partial absorption cross-section $\sigma_\ell$ for a particular $\ell$ is given by \cite{Dey:2018cws},
\begin{equation}
    \sigma_\ell = \frac{\pi (2\ell + 1)}{\omega^2} |T_\ell(\omega)|^2
\end{equation}
Total absorption cross-section is defined as,
\begin{equation}
    \sigma=\sum_\ell \sigma_\ell =\sum_\ell \frac{\pi (2\ell + 1)}{\omega^2} |T_\ell(\omega)|^2 .
\end{equation}

\begin{figure}[t]
\centering
\includegraphics[width=\textwidth]{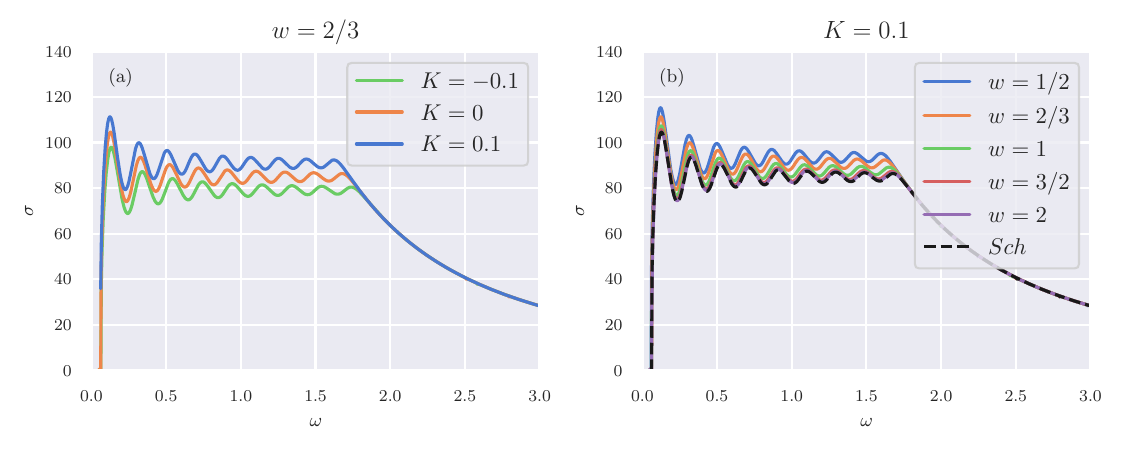}
\caption{Effect of \( K \) (left) and \( w \) (right) on total absorption cross section $\sigma$ vs QNM frequency $\omega$ for scalar perturbation. To find $\sigma$ we have added up to 10 modes from $\ell=0$ to $\ell=10$} 
\label{fig9}
\end{figure}

The variation of $\sigma$ with anisotropic matter field parameters is studied in Fig. \ref{fig9} for scalar perturbations. A similar behavior is observed for electromagnetic perturbations. From the figure, three distinct regions can be identified. The first region shows an initial increase in $\sigma$, attributed to the rise in the transmission coefficient \( |T(\omega)| \) with $\omega$. In the second region, $\sigma$ exhibits oscillatory behavior, arising from contributions of different $\ell$ modes. Finally, in the power-law fall-off region, $\sigma$ decreases due to the saturation of \( |T(\omega)| \) to 1 at higher frequencies, as observed in Fig. \ref{greybody}. Beyond this, as $\omega$ continues to increase, $\sigma$ becomes proportional to $1/\omega^2$, regardless of the values of the anisotropic parameters. The effect of the parameter $K$ is analyzed in \ref{fig9}(a), where splitting behavior is observed. The strength of this splitting depends on the values of $w$, as depicted in \ref{fig9}(b).

\section{Results and Discussion}\label{sec_disc}
In this study, we explored the perturbations of black holes surrounded by an anisotropic matter field, focusing on a family of solutions that generalize the Reissner-Nordström spacetime. These black holes possess additional characteristics, or ``hair,'' due to the negative radial pressure of the surrounding anisotropic matter. Astrophysical black holes are seldom isolated and often coexist with matter or fields, it becomes essential to understand how such matter or fields influence black hole properties and observable phenomena. We investigated massless scalar and electromagnetic field perturbations in the background of these black holes. Employing the optimal order WKB method, we calculated the quasinormal modes (QNMs), which describe the characteristic oscillations of black holes in response to perturbations. First we derive the effective potentials for these perturbations, and then we assess how the anisotropic matter field affects the oscillation frequencies associated with QNMs. The error estimation confirms the reliability of the computed QNM frequencies. For completeness we also comment on gravitational perturbations in the appendix \ref{gwapp}.

Our findings reveal that the presence of an anisotropic matter field leads to a splitting in the QNM frequencies when compared to the Schwarzschild black hole case. Specifically, for positive values of the anisotropy parameter $K$, the real part of the QNM frequencies decreases, whereas for negative values, it increases. This splitting is consistent across different angular momentum modes and becomes more pronounced for smaller values of the anisotropy parameter \( w \), indicating that less anisotropic matter exerts a greater influence on the QNMs.

We have also examined the critical orbits around the black hole to investigate the connection between the eikonal QNMs, the shadow radius of the black hole, and the Lyapunov exponent, which characterizes the instability timescale of photon sphere orbits. We observe that the shadow radius increases with positive anisotropy parameter $K$ values and decreases with negative values, mirroring the behavior of the real part of the QNM frequencies. Similarly, the Lyapunov exponent exhibits dependence on the anisotropy parameters, aligning with the imaginary part of the QNMs. The correlation between shadow and QNM suggest that the anisotropic matter field has a significant impact on the observable features of black holes, such as their shadows.

Additionally, we analyzed the scattering properties and grey-body factors resulting from the perturbations. The anisotropic matter field alters the potential barrier surrounding the black hole, influencing the transmission and reflection coefficients of waves propagating in this spacetime. Our results indicate that the grey-body factor increases with larger positive values of the anisotropy parameter $K$, implying that the potential barrier becomes more penetrable and allows more radiation to escape to infinity. This effect is more pronounced for lower values of \( w \), which is consistent with the trends observed in the QNM frequencies and effective potentials.

Our results emphasize that increasing the value of the anisotropy parameter \( w \) causes the black hole's properties to converge toward those of the Schwarzschild case. This indicates that the influence of the anisotropic matter field diminishes with higher \( w \) values, and the black hole behaves more like an isolated Schwarzschild black hole.

Considering that real astrophysical black holes are often rotating and surrounded by matter fields, future research could extend this analysis to rotating black holes immersed in anisotropic matter. Additionally, exploring the perturbations and quasinormal modes of wormhole solutions with anisotropic matter fields could provide further insights into distinguishing these exotic objects from black holes.

\appendix
\section{Error Calculation} \label{errapp}

In this appendix, we provide the error estimation in the WKB approximation.

\begin{table}[h!]
\centering
\begin{tabular}{m{6em} m{12em} m{7em} m{12em} m{6em}}
\hline
WKB Order & Scalar QNM (\( \ell = 0 \)) & Error & EM QNM (\( \ell = 1 \)) & Error\\
\hline
 12 & 0.070246 - 0.302642 i & 0.234244 & 1.709442 + 210.9719 i & 2680.561 \\ 
 11 & 0.138488 - 0.153511 i & 0.088657 & 19.03589 + 18.94548 i & 105.4897 \\ 
 10 & 0.112319 - 0.130392 i & 0.020183 & 1.853118 - 0.007509 i & 13.36668 \\ 
 9  & 0.127818 - 0.114581 i & 0.014954 & 0.234561 - 0.059325 i & 0.805055 \\ 
 8  & 0.115497 - 0.100653 i & 0.009595 & 0.245283 - 0.093083 i & 0.017688 \\ 
 7  & 0.111850 - 0.103934 i & 0.003016 & 0.246318 - 0.092692 i & 0.000558 \\ 
 6  & 0.109512 - 0.101414 i & 0.003911 & 0.246374 - 0.092840 i & 0.000088 \\ 
 5  & 0.104414 - 0.106365 i & 0.004626 & 0.246449 - 0.092811 i & 0.000134 \\ 
 4  & 0.108763 - 0.110637 i & 0.004874 & 0.246359 - 0.092572 i & 0.001244 \\ 
 3  & 0.103659 - 0.116083 i & 0.019581 & 0.244044 - 0.093450 i & 0.006366 \\ 
\hline
\end{tabular}
\caption{Error estimation in the WKB method for scalar and EM perturbations with \( n=0 \), \( M=1 \), \( Q=0 \), \( K=0.1 \) and \( w = 3/2 \). We have performed similar calculations in every case for both scalar and electromagnetic perturbations.} 
\label{tabapp}
\end{table}

\begin{figure}[h]
\centering
\includegraphics[width=\textwidth]{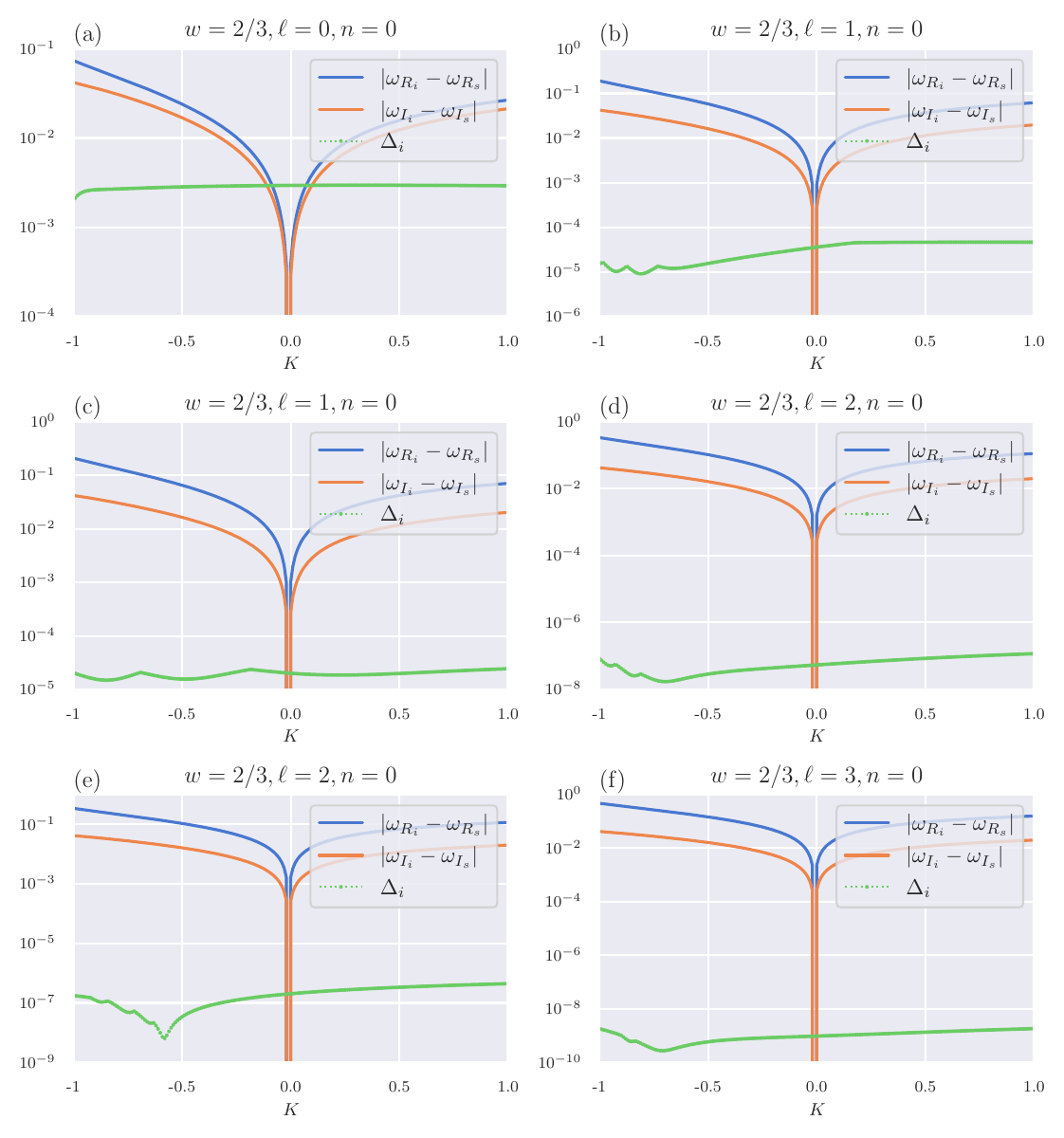}
\caption{Error estimation in optimal order WKB approximation compared with the effect of anisotropic matter field.} 
\label{fig3}
\end{figure}

\section{Gravitational perturbation}\label{gwapp}
In this appendix, we briefly present the details of the gravitational perturbation. The perturbation and quasinormal mode calculation of black hole solutions with non-vanishing energy-momentum tensor have been studied in the literature by neglecting the perturbations of the energy-momentum tensor \cite{Zhang:2006ij, Konoplya:2006ar, Ding:2018nhz, Yang:2022ifo}. That is, instead of considering the full perturbation equations
\begin{equation}
\delta (R_{\mu \nu} - \frac{1}{2} R g_{\mu \nu}) = \kappa \delta T_{\mu \nu},
\end{equation}
only the reduced perturbation equations were considered,
\begin{equation}\label{reduced}
\delta (R_{\mu \nu} - \frac{1}{2} R g_{\mu \nu}) = 0.
\end{equation}
However, in this appendix, we present the full perturbation of the Einstein equation, as neglecting the perturbation of the energy-momentum tensor is not always justified \cite{Churilova:2020bql}.

The gravitational perturbation of a spherically symmetric black hole spacetime can be studied using the Regge-Wheeler/Zerilli formalism \cite{Regge:1957td, Edelstein:1970sk, Zerilli:1970se}. Considering small perturbations to the background metric $g^{(0)}_{\mu\nu}$ in the region outside the black hole,
\begin{equation}
    g_{\mu\nu}=g_{\mu\nu}^{\left(0\right)}+\varepsilon h_{\mu\nu},
    \label{lin_pert}
\end{equation}
where the perturbation is represented by a symmetric tensor $h_{\mu \nu}$. In the linear order, the perturbed Ricci tensor, Ricci scalar, and energy-momentum tensor are given by,
\begin{equation}
    R_{\mu\nu}=R_{\mu\nu}^{\left(0\right)}+\varepsilon R_{\mu\nu}^{\left(1\right)} \, ,
\qquad
    R=R^{\left(0\right)}+\varepsilon R^{\left(1\right)} \, ,
\qquad
    T_{\mu\nu}=T_{\mu\nu}^{\left(0\right)}+\varepsilon T_{\mu\nu}^{\left(1\right)} \, ,
\end{equation}
The linear-order perturbation correction for the Einstein equation becomes,
\begin{equation}
    R_{\mu\nu}^{\left(1\right)}-\frac{1}{2}\left[g_{\mu\nu}^{\left(0\right)}R^{\left(1\right)}+h_{\mu\nu}R^{\left(0\right)}\right]=8\pi T_{\mu\nu}^{\left(1\right)} \, ,
    \label{pert_eqn}
\end{equation}
where the superscripts represent the order of perturbation.

In the Regge-Wheeler formalism, general spherically symmetric tensorial perturbations $h_{\mu\nu}$ are split into odd wave (axial, Regge-Wheeler or vector-type perturbations) and even wave (polar, Zerilli or scalar-type perturbations) with parities equal to $(-1)^{\ell+1}$ and $(-1)^\ell$, respectively \cite{Nagar:2005ea,Thorne:1980ru,Zerilli:1970se}. In the linear order, the odd and even perturbations decouple, and they can be treated separately. Due to the principle of general covariance, the theory should be covariant under infinitesimal coordinate transformations. Thus, we can choose a specific gauge to simplify the problem, such as the Regge-Wheeler gauge \cite{Regge:1957td,Vishveshwara:1970cc}. Due to the static nature of the spacetime, the time dependence can be decomposed as $h_{\mu \nu} (x,t)=e^{-i\omega t} h_{\mu \nu}(x)$, where $\omega$ is the quasinormal frequency of perturbations. Due to spherical symmetry, one need not work with arbitrary $m$ and therefore, without loss of generality, we can set $m=0$. The odd wave perturbations in Regge-Wheeler canonical gauge are given by \cite{Regge:1957td},
\begin{equation}
    \left(h_{\mu\nu}\right)_{\text{odd}} =\begin{pmatrix}
    {0}&{0}&{0}&{h_{0}}\\
    {0}&{0}&{0}&{h_{1}}\\
    {0}&{0}&{0}&{0}\\
    {h_{0}}&{h_{1}}&{0}&{0}\\
    \end{pmatrix}e^{-i\omega t}\sin\theta  P_\ell'\left(\cos\theta\right)
    \label{odd_decomp}
\end{equation}
In the canonical gauge, the odd perturbation of the energy-momentum tensor is given by \cite{Nagar:2005ea, Zerilli:1974ai},
\begin{equation}
 \left(T_{\mu\nu}^{\left(1\right)}\right)_{\text{odd}} = e^{-i\omega t} \sin\theta \left[ P_\ell\left(\cos\theta\right) \, \mathbf{A} + P_\ell'(\cos\theta) \, \mathbf{B} \right]  
\label{em_decomp}
\end{equation}
where
\[
\mathbf{A} =
\begin{pmatrix}
0 & 0 & 0 & 0 \\
0 & 0 & 0 & 0 \\
0 & 0 & 0 & \dfrac{t_2 \ell(\ell+1)}{2} \\
0 & 0 & \text{sym} & 0
\end{pmatrix},
\qquad
\mathbf{B} =\begin{pmatrix}
0 & 0 & 0 & t_0 \\
0 & 0 & 0 & t_1 \\
0 & 0 & 0 & t_2 \cot\theta \\
\text{sym} & \text{sym} & \text{sym} & 0
\end{pmatrix}.
\]
where $h_0$, $h_1$ are functions of the radial coordinate, and $t_0$, $t_1$, $t_2$ are the radial functions that determine the perturbation of the energy-momentum tensor.

Substituting \eqref{odd_decomp} and \eqref{em_decomp} into \eqref{pert_eqn}, we get three non-trivial equations. From the $(t, \theta)$ component we get,
\begin{equation}
    \frac{h_0 \left[r^2 f''+2 r f'+2 f+\ell(\ell+1)-2\right]-i r f \left[2 \omega  h_1+r \omega  h_1'-i r h_0''\right]}{2 r^2}=8\pi  t_0  .
    \label{03eqn}
\end{equation}
From the $(r, \theta)$ component we get,
\begin{equation}
    \frac{h_1 \left[f \left(r^2 f''+2 r f'+\ell(\ell+1)-2\right)-r^2 \omega ^2\right]+i r^2 \omega  h_0'-2 i r \omega  h_0}{2 r^2 f}=8 \pi  t_1 .
    \label{13eqn}
\end{equation}
From the $(\theta , \varphi)$ component we get,
\begin{equation}
    \frac{i \omega  h_0}{f}+h_1 f'+f h_1'=-8 \pi  t_2 .
    \label{23eqn}
\end{equation}
There are three equations with two unknowns, $h_0$ and $h_1$, therefore not all these equations are independent, so we need to consider only two equations. First, we solve \eqref{23eqn} for $h_0$ in terms of $h_1$ and $t_2$,
\begin{equation}
    h_0=\frac{i f}{\omega } \left[h_1 f'+f h_1'+8 \pi  t_2\right]
    \label{value of h0(r)}
\end{equation}
and then substitute it into \eqref{13eqn} to get, 
\begin{multline}
    h_1'' =\frac{1}{r^2 f^2} \Big[h_1 \left(f \left(4 r f'+\ell^2+\ell-2\right)-r^2 \left(f'^2+\omega ^2\right)\right)\\
     -r \left(f \left(3 r f' h_1'+16 \pi  r t_1+8 \pi  r t_2'-16 \pi  t_2\right)
     +8 \pi  r t_2 f'-2 f^2 h_1'\right)\Big] . 
     \label{diffh1}
\end{multline}
Redefining the function $h_1$ as,
\begin{equation}
    h_1 (r) = \frac{r}{f} \psi{\left(r \right)}
\end{equation}
and using the tortoise coordinates given in \eqref{tortoise_def}, the equation \eqref{diffh1} takes the Schr\"{o}dinger-like form,
\begin{equation}
    \frac{d^{2}\psi{\left(r_{\star} \right)}}{d r_{\star}^{2}}+ \left[\omega ^2-V_{{RW}}\right]\psi (r_\star)=S_{\text{odd}} \, , 
    \label{odd_sch}
\end{equation}
where the source term $S_{\text{odd}}(r)$ is given by,
\begin{equation}
    S_{\text{odd}}= \frac{8 \pi  f}{r^2}  \left[2 f \left(t_2-r t_1\right)-r \left(f' t_2+t_2'\right)\right] .
\end{equation}
In the above expression, $V_{{RW}}$ is the Regge-Wheeler potential for odd perturbations,
\begin{equation}
     V_{RW}(r)=f(r) \left(\frac{\ell\left(\ell+1\right)+2 \left[f(r)-1\right]}{r^{2}}+\frac{f^{\prime}(r)}{r}+f^{\prime\prime}(r)\right) .
    \label{Odd Effective Potential}
\end{equation}
For the metric function \eqref{metric_function}, it becomes,
\begin{equation}
     V_{RW}(r) = \left(1-\frac{2M}{r} +\frac{Q^2}{r^2} - \frac{K}{r^{2w}}\right) \left(\frac{\ell (\ell+1)}{r^2}-\frac{6 M}{r^3}+\frac{6 Q^2}{r^4}-\frac{2 K \left(1+2 w^2\right)}{r^{2 (w+1)}} \right) .
\end{equation}
The equation \eqref{odd_sch} describes the evolution of perturbations of black holes surrounded by an anisotropic matter field, wherein the perturbation has mass-energy much smaller than that of the black hole. The solution of the evolution equation with appropriate boundary conditions governs the reaction of the black hole to the perturbation, manifesting as the emission of gravitational radiation. For the vanishing source term in \eqref{odd_sch}, the solutions $\omega$ are referred to as quasinormal modes \cite{Nagar:2005ea}.

For completeness, we provide essential equations in even wave perturbations in Regge-Wheeler canonical gauge (also known as Zerilli gauge \cite{Zerilli:1970se}). The perturbation of the metric is decomposed as,
\begin{equation}
\left(h_{\mu\nu}\right)_{\text{Even}}=\begin{pmatrix}
{fh_{0}}&{h_{1}}&{0}&{0}\\
{h_{1}}&\displaystyle\frac{H_{2}\left(r\right)}{f}&{0}&{0}\\
{0}&{0}&{r^{2}K\left(r\right)}&{0}\\
{0}&{0}&{0}&{r^{2}\sin^{2}\left(\theta\right)K\left(r\right)}\\
\end{pmatrix}e^{-i\omega t}P_{\ell}\left(\cos\left(\theta\right)\right)
\end{equation}
The even perturbation of the energy-momentum tensor is decomposed as,
\begin{equation}
T^{(1)}_{\text{Even}} = e^{-i\omega t} \left[ P_\ell(\cos\theta) \, \mathbf{C}(r) + P_\ell'\left(\cos\theta\right) \, \mathbf{D}(r) \right]    
\end{equation}
where
\[
\mathbf{C}(r) =
\begin{pmatrix}
f T_0(r) & T_1(r) & 0 & 0 \\
\text{sym} & \dfrac{T_2(r)}{f} & 0 & 0 \\
0 & 0 & -r^2 \left[ \ell(\ell+1) T_6(r) - T_3(r) \right] & 0 \\
0 & 0 & 0 & r^2 \sin^2\theta \, T_3(r)
\end{pmatrix},
\]
\[
\mathbf{D}(r) =
\begin{pmatrix}
0 & 0 & T_4(r) & 0 \\
0 & 0 & T_5(r) & 0 \\
\text{sym} & \text{sym} & -r^2 \cot\theta \, T_6(r) & 0 \\
0 & 0 & 0 & r^2 \sin\theta \cos\theta \, T_6(r)
\end{pmatrix}.
\]
where $H_{0}$, $H_{1}$, $K$ and $T_0$, $T_1$, $T_2$, $T_3$, $T_4$, $T_5$, $T_6$ are functions of the radial coordinate. Following \cite{Zerilli:1970se}, we obtain the Zerilli potential:
\begin{multline}
    V_{\text{even}}\left(r\right)=\frac{f}{r^2 \left(3 r f'+2 \rho \right)^2} \Big[\left(3 r f'+2 \rho \right) \left(r \left(r f'' \left(\rho -3 r f'\right)-3 f' \left(r f'-2 \rho \right)\right)+4 \rho ^2\right)\\
     +f \left(3 r \left(3 r^3 f''\,^2-2 r \rho  f''+8 \rho  f'+6 r f'^2-r^2 \left(3 r f'+2 \rho \right)f''' \right)+8 \rho ^2\right)\Big]
\end{multline}
where $\rho$ is given by the relation $\rho=\frac{1}{2}\left(\ell-1\right)\left(\ell+2\right)$.

\bibliography{BibTex}

\providecommand{\href}[2]{#2}\begingroup\raggedright\begin{thebibliography}{10}

\bibitem{Abbott:2016blz}
{\scshape LIGO Scientific, Virgo} collaboration, \emph{{Observation of Gravitational Waves from a Binary Black Hole Merger}}, \href{https://doi.org/10.1103/PhysRevLett.116.061102}{\emph{Phys. Rev. Lett.} {\bfseries 116} (2016) 061102} [\href{https://arxiv.org/abs/1602.03837}{{\ttfamily 1602.03837}}].

\bibitem{TheLIGOScientific:2016src}
{\scshape LIGO Scientific, Virgo} collaboration, \emph{{Tests of general relativity with GW150914}}, \href{https://doi.org/10.1103/PhysRevLett.116.221101}{\emph{Phys. Rev. Lett.} {\bfseries 116} (2016) 221101} [\href{https://arxiv.org/abs/1602.03841}{{\ttfamily 1602.03841}}].

\bibitem{Abbott:2016nmj}
{\scshape LIGO Scientific, Virgo} collaboration, \emph{{GW151226: Observation of Gravitational Waves from a 22-Solar-Mass Binary Black Hole Coalescence}}, \href{https://doi.org/10.1103/PhysRevLett.116.241103}{\emph{Phys. Rev. Lett.} {\bfseries 116} (2016) 241103} [\href{https://arxiv.org/abs/1606.04855}{{\ttfamily 1606.04855}}].

\bibitem{Akiyama:2019bqs}
{\scshape Event Horizon Telescope} collaboration, \emph{{First M87 Event Horizon Telescope Results. IV. Imaging the Central Supermassive Black Hole}}, \href{https://doi.org/10.3847/2041-8213/ab0e85}{\emph{Astrophys. J. Lett.} {\bfseries 875} (2019) L4} [\href{https://arxiv.org/abs/1906.11241}{{\ttfamily 1906.11241}}].

\bibitem{Akiyama:2019cqa}
{\scshape Event Horizon Telescope} collaboration, \emph{{First M87 Event Horizon Telescope Results. I. The Shadow of the Supermassive Black Hole}}, \href{https://doi.org/10.3847/2041-8213/ab0ec7}{\emph{Astrophys. J.} {\bfseries 875} (2019) L1} [\href{https://arxiv.org/abs/1906.11238}{{\ttfamily 1906.11238}}].

\bibitem{Akiyama:2019fyp}
{\scshape Event Horizon Telescope} collaboration, \emph{{First M87 Event Horizon Telescope Results. V. Physical Origin of the Asymmetric Ring}}, \href{https://doi.org/10.3847/2041-8213/ab0f43}{\emph{Astrophys. J. Lett.} {\bfseries 875} (2019) L5} [\href{https://arxiv.org/abs/1906.11242}{{\ttfamily 1906.11242}}].

\bibitem{Regge:1957td}
T.~Regge and J.~A. Wheeler, \emph{{Stability of a Schwarzschild singularity}}, \href{https://doi.org/10.1103/PhysRev.108.1063}{\emph{Phys. Rev.} {\bfseries 108} (1957) 1063}.

\bibitem{Edelstein:1970sk}
L.~A. Edelstein and C.~V. Vishveshwara, \emph{{Differential equations for perturbations on the schwarzschild metric}}, \href{https://doi.org/10.1103/PhysRevD.1.3514}{\emph{Phys. Rev. D} {\bfseries 1} (1970) 3514}.

\bibitem{Zerilli:1970se}
F.~J. Zerilli, \emph{{Effective potential for even parity Regge-Wheeler gravitational perturbation equations}}, \href{https://doi.org/10.1103/PhysRevLett.24.737}{\emph{Phys. Rev. Lett.} {\bfseries 24} (1970) 737}.

\bibitem{Castellani:2019pvh}
L.~Castellani, A.~Ceresole, R.~D'Auria and P.~Fr\'e, eds., \emph{{Tullio Regge: An Eclectic Genius}: {~From Quantum Gravity to Computer Play}}. World Scientific, 9, 2019, \href{https://doi.org/10.1142/11643}{10.1142/11643}.

\bibitem{Price:1994pm}
R.~H. Price and J.~Pullin, \emph{{Colliding black holes: The Close limit}}, \href{https://doi.org/10.1103/PhysRevLett.72.3297}{\emph{Phys. Rev. Lett.} {\bfseries 72} (1994) 3297} [\href{https://arxiv.org/abs/gr-qc/9402039}{{\ttfamily gr-qc/9402039}}].

\bibitem{Abrahams:1994qu}
A.~M. Abrahams and G.~B. Cook, \emph{{Collisions of boosted black holes: perturbation theory prediction of gravitational radiation}}, \href{https://doi.org/10.1103/PhysRevD.50.R2364}{\emph{Phys. Rev. D} {\bfseries 50} (1994) R2364} [\href{https://arxiv.org/abs/gr-qc/9405051}{{\ttfamily gr-qc/9405051}}].

\bibitem{Abrahams:1994xy}
A.~M. Abrahams, S.~L. Shapiro and S.~A. Teukolsky, \emph{{Calculation of gravitational wave forms from black hole collisions and disk collapse: Applying perturbation theory to numerical space-times}}, \href{https://doi.org/10.1103/PhysRevD.51.4295}{\emph{Phys. Rev. D} {\bfseries 51} (1995) 4295} [\href{https://arxiv.org/abs/gr-qc/9408036}{{\ttfamily gr-qc/9408036}}].

\bibitem{Kokkotas:1999bd}
K.~D. Kokkotas and B.~G. Schmidt, \emph{{Quasinormal modes of stars and black holes}}, \href{https://doi.org/10.12942/lrr-1999-2}{\emph{Living Rev. Rel.} {\bfseries 2} (1999) 2} [\href{https://arxiv.org/abs/gr-qc/9909058}{{\ttfamily gr-qc/9909058}}].

\bibitem{Berti:2009kk}
E.~Berti, V.~Cardoso and A.~O. Starinets, \emph{{Quasinormal modes of black holes and black branes}}, \href{https://doi.org/10.1088/0264-9381/26/16/163001}{\emph{Class. Quant. Grav.} {\bfseries 26} (2009) 163001} [\href{https://arxiv.org/abs/0905.2975}{{\ttfamily 0905.2975}}].

\bibitem{Konoplya:2011qq}
R.~A. Konoplya and A.~Zhidenko, \emph{{Quasinormal modes of black holes: From astrophysics to string theory}}, \href{https://doi.org/10.1103/RevModPhys.83.793}{\emph{Rev. Mod. Phys.} {\bfseries 83} (2011) 793} [\href{https://arxiv.org/abs/1102.4014}{{\ttfamily 1102.4014}}].

\bibitem{Chandrasekhar:1975zza}
S.~Chandrasekhar and S.~L. Detweiler, \emph{{The quasi-normal modes of the Schwarzschild black hole}}, \href{https://doi.org/10.1098/rspa.1975.0112}{\emph{Proc. Roy. Soc. Lond. A} {\bfseries 344} (1975) 441}.

\bibitem{Vishveshwara:1970zz}
C.~V. Vishveshwara, \emph{{Scattering of Gravitational Radiation by a Schwarzschild Black-hole}}, \href{https://doi.org/10.1038/227936a0}{\emph{Nature} {\bfseries 227} (1970) 936}.

\bibitem{Ching:1998mxl}
E.~S.~C. Ching, P.~T. Leung, A.~Maassen van~den Brink, W.~M. Suen, S.~S. Tong and K.~Young, \emph{{Quasinormal-mode expansion for waves in open systems}}, \href{https://doi.org/10.1103/RevModPhys.70.1545}{\emph{Rev. Mod. Phys.} {\bfseries 70} (1998) 1545} [\href{https://arxiv.org/abs/gr-qc/9904017}{{\ttfamily gr-qc/9904017}}].

\bibitem{Nollert:1998ys}
H.-P. Nollert and R.~H. Price, \emph{{Quantifying excitations of quasinormal mode systems}}, \href{https://doi.org/10.1063/1.532698}{\emph{J. Math. Phys.} {\bfseries 40} (1999) 980} [\href{https://arxiv.org/abs/gr-qc/9810074}{{\ttfamily gr-qc/9810074}}].

\bibitem{Nollert:1999ji}
H.-P. Nollert, \emph{{TOPICAL REVIEW: Quasinormal modes: the characteristic `sound' of black holes and neutron stars}}, \href{https://doi.org/10.1088/0264-9381/16/12/201}{\emph{Class. Quant. Grav.} {\bfseries 16} (1999) R159}.

\bibitem{Nagar:2005ea}
A.~Nagar and L.~Rezzolla, \emph{{Gauge-invariant non-spherical metric perturbations of Schwarzschild black-hole spacetimes}}, \href{https://doi.org/10.1088/0264-9381/22/16/R01}{\emph{Class. Quant. Grav.} {\bfseries 22} (2005) R167} [\href{https://arxiv.org/abs/gr-qc/0502064}{{\ttfamily gr-qc/0502064}}].

\bibitem{Ferrari:2007dd}
V.~Ferrari and L.~Gualtieri, \emph{{Quasi-Normal Modes and Gravitational Wave Astronomy}}, \href{https://doi.org/10.1007/s10714-007-0585-1}{\emph{Gen. Rel. Grav.} {\bfseries 40} (2008) 945} [\href{https://arxiv.org/abs/0709.0657}{{\ttfamily 0709.0657}}].

\bibitem{Stephani:2003tm}
H.~Stephani, D.~Kramer, M.~A.~H. MacCallum, C.~Hoenselaers and E.~Herlt, \emph{{Exact solutions of Einstein's field equations}}, Cambridge Monographs on Mathematical Physics. Cambridge Univ. Press, Cambridge, 2003, \href{https://doi.org/10.1017/CBO9780511535185}{10.1017/CBO9780511535185}.

\bibitem{Delgaty:1998uy}
M.~S.~R. Delgaty and K.~Lake, \emph{{Physical acceptability of isolated, static, spherically symmetric, perfect fluid solutions of Einstein's equations}}, \href{https://doi.org/10.1016/S0010-4655(98)00130-1}{\emph{Comput. Phys. Commun.} {\bfseries 115} (1998) 395} [\href{https://arxiv.org/abs/gr-qc/9809013}{{\ttfamily gr-qc/9809013}}].

\bibitem{Kim:2019hfp}
H.-C. Kim, B.-H. Lee, W.~Lee and Y.~Lee, \emph{{Rotating black holes with an anisotropic matter field}}, \href{https://doi.org/10.1103/PhysRevD.101.064067}{\emph{Phys. Rev. D} {\bfseries 101} (2020) 064067} [\href{https://arxiv.org/abs/1912.09709}{{\ttfamily 1912.09709}}].

\bibitem{Cho:2017nhx}
I.~Cho and H.-C. Kim, \emph{{Simple black holes with anisotropic fluid}}, \href{https://doi.org/10.1088/1674-1137/43/2/025101}{\emph{Chin. Phys. C} {\bfseries 43} (2019) 025101} [\href{https://arxiv.org/abs/1703.01103}{{\ttfamily 1703.01103}}].

\bibitem{Badia:2020pnh}
J.~Badía and E.~F. Eiroa, \emph{{Influence of an anisotropic matter field on the shadow of a rotating black hole}}, \href{https://doi.org/10.1103/PhysRevD.102.024066}{\emph{Phys. Rev. D} {\bfseries 102} (2020) 024066} [\href{https://arxiv.org/abs/2005.03690}{{\ttfamily 2005.03690}}].

\bibitem{AhmedRizwan:2020sza}
C.~L. Ahmed~Rizwan, A.~Naveena~Kumara, K.~Hegde, M.~S. Ali and K.~M. Ajith, \emph{{Rotating black hole with an anisotropic matter field as a particle accelerator}}, \href{https://doi.org/10.1088/1361-6382/abe2d9}{\emph{Class. Quant. Grav.} {\bfseries 38} (2021) 075030} [\href{https://arxiv.org/abs/2008.01426}{{\ttfamily 2008.01426}}].

\bibitem{Kim:2019ojs}
H.-C. Kim and Y.~Lee, \emph{{Spherically Symmetric Wormholes with anisotropic matter}}, \href{https://doi.org/10.1088/1475-7516/2019/09/001}{\emph{JCAP} {\bfseries 09} (2019) 001} [\href{https://arxiv.org/abs/1905.10050}{{\ttfamily 1905.10050}}].

\bibitem{Gibbons:1976ue}
G.~W. Gibbons and S.~W. Hawking, \emph{{Action Integrals and Partition Functions in Quantum Gravity}}, \href{https://doi.org/10.1103/PhysRevD.15.2752}{\emph{Phys. Rev. D} {\bfseries 15} (1977) 2752}.

\bibitem{Hawking:1995ap}
S.~W. Hawking and S.~F. Ross, \emph{{Duality between electric and magnetic black holes}}, \href{https://doi.org/10.1103/PhysRevD.52.5865}{\emph{Phys. Rev. D} {\bfseries 52} (1995) 5865} [\href{https://arxiv.org/abs/hep-th/9504019}{{\ttfamily hep-th/9504019}}].

\bibitem{Kiselev:2002dx}
V.~Kiselev, \emph{{Quintessence and black holes}}, \href{https://doi.org/10.1088/0264-9381/20/6/310}{\emph{Class. Quant. Grav.} {\bfseries 20} (2003) 1187} [\href{https://arxiv.org/abs/gr-qc/0210040}{{\ttfamily gr-qc/0210040}}].

\bibitem{Zwicky:1933gu}
F.~Zwicky, \emph{{Die Rotverschiebung von extragalaktischen Nebeln}}, \href{https://doi.org/10.1007/s10714-008-0707-4}{\emph{Helv. Phys. Acta} {\bfseries 6} (1933) 110}.

\bibitem{Rubin:1970zza}
V.~C. Rubin and W.~K. Ford, Jr., \emph{{Rotation of the Andromeda Nebula from a Spectroscopic Survey of Emission Regions}}, \href{https://doi.org/10.1086/150317}{\emph{Astrophys. J.} {\bfseries 159} (1970) 379}.

\bibitem{Zou:2019ays}
D.-C. Zou and Y.~S. Myung, \emph{{Scalar hairy black holes in Einstein-Maxwell-conformally coupled scalar theory}}, \href{https://doi.org/10.1016/j.physletb.2020.135332}{\emph{Phys. Lett. B} {\bfseries 803} (2020) 135332} [\href{https://arxiv.org/abs/1911.08062}{{\ttfamily 1911.08062}}].

\bibitem{Lee:2021sws}
B.-H. Lee, W.~Lee and Y.~S. Myung, \emph{{Shadow cast by a rotating black hole with anisotropic matter}}, \href{https://doi.org/10.1103/PhysRevD.103.064026}{\emph{Phys. Rev. D} {\bfseries 103} (2021) 064026} [\href{https://arxiv.org/abs/2101.04862}{{\ttfamily 2101.04862}}].

\bibitem{Chen:2005qh}
S.-b. Chen and J.-l. Jing, \emph{{Quasinormal modes of a black hole surrounded by quintessence}}, \href{https://doi.org/10.1088/0264-9381/22/21/011}{\emph{Class. Quant. Grav.} {\bfseries 22} (2005) 4651} [\href{https://arxiv.org/abs/gr-qc/0511085}{{\ttfamily gr-qc/0511085}}].

\bibitem{Cardoso:2001bb}
V.~Cardoso and J.~P.~S. Lemos, \emph{{Quasinormal modes of Schwarzschild anti-de Sitter black holes: Electromagnetic and gravitational perturbations}}, \href{https://doi.org/10.1103/PhysRevD.64.084017}{\emph{Phys. Rev. D} {\bfseries 64} (2001) 084017} [\href{https://arxiv.org/abs/gr-qc/0105103}{{\ttfamily gr-qc/0105103}}].

\bibitem{Dey:2018cws}
S.~Dey and S.~Chakrabarti, \emph{{A note on electromagnetic and gravitational perturbations of the Bardeen de Sitter black hole: quasinormal modes and greybody factors}}, \href{https://doi.org/10.1140/epjc/s10052-019-7004-0}{\emph{Eur. Phys. J. C} {\bfseries 79} (2019) 504} [\href{https://arxiv.org/abs/1807.09065}{{\ttfamily 1807.09065}}].

\bibitem{Chandrasekhar:1985kt}
S.~Chandrasekhar, \emph{{The mathematical theory of black holes}}. 1985.

\bibitem{Schutz:1985km}
B.~F. Schutz and C.~M. Will, \emph{{Black hole normal modes - A semianalytic approach}}, \href{https://doi.org/10.1086/184453}{\emph{Astrophys. J. Lett.} {\bfseries 291} (1985) L33}.

\bibitem{Iyer:1986np}
S.~Iyer and C.~M. Will, \emph{{Black Hole Normal Modes: A {WKB} Approach. 1. Foundations and Application of a Higher Order {WKB} Analysis of Potential Barrier Scattering}}, \href{https://doi.org/10.1103/PhysRevD.35.3621}{\emph{Phys. Rev. D} {\bfseries 35} (1987) 3621}.

\bibitem{Konoplya:2003ii}
R.~A. Konoplya, \emph{{Quasinormal behavior of the d-dimensional Schwarzschild black hole and higher order WKB approach}}, \href{https://doi.org/10.1103/PhysRevD.68.024018}{\emph{Phys. Rev. D} {\bfseries 68} (2003) 024018} [\href{https://arxiv.org/abs/gr-qc/0303052}{{\ttfamily gr-qc/0303052}}].

\bibitem{Matyjasek:2017psv}
J.~Matyjasek and M.~Opala, \emph{{Quasinormal modes of black holes. The improved semianalytic approach}}, \href{https://doi.org/10.1103/PhysRevD.96.024011}{\emph{Phys. Rev. D} {\bfseries 96} (2017) 024011} [\href{https://arxiv.org/abs/1704.00361}{{\ttfamily 1704.00361}}].

\bibitem{Konoplya:2019hlu}
R.~A. Konoplya, A.~Zhidenko and A.~F. Zinhailo, \emph{{Higher order WKB formula for quasinormal modes and grey-body factors: recipes for quick and accurate calculations}}, \href{https://doi.org/10.1088/1361-6382/ab2e25}{\emph{Class. Quant. Grav.} {\bfseries 36} (2019) 155002} [\href{https://arxiv.org/abs/1904.10333}{{\ttfamily 1904.10333}}].

\bibitem{Leaver:1985ax}
E.~W. Leaver, \emph{{An Analytic representation for the quasi normal modes of Kerr black holes}}, \href{https://doi.org/10.1098/rspa.1985.0119}{\emph{Proc. Roy. Soc. Lond. A} {\bfseries 402} (1985) 285}.

\bibitem{Ciric:2017rnf}
M.~D. \'Ciri\'c, N.~Konjik and A.~Samsarov, \emph{{Noncommutative scalar quasinormal modes of the Reissner\textendash{}Nordstr\"om black hole}}, \href{https://doi.org/10.1088/1361-6382/aad201}{\emph{Class. Quant. Grav.} {\bfseries 35} (2018) 175005} [\href{https://arxiv.org/abs/1708.04066}{{\ttfamily 1708.04066}}].

\bibitem{DimitrijevicCiric:2019hqq}
M.~Dimitrijevi\'c~\'Ciri\'c, N.~Konjik and A.~Samsarov, \emph{{Noncommutative scalar field in the nonextremal Reissner-Nordstr\"om background: Quasinormal mode spectrum}}, \href{https://doi.org/10.1103/PhysRevD.101.116009}{\emph{Phys. Rev. D} {\bfseries 101} (2020) 116009} [\href{https://arxiv.org/abs/1904.04053}{{\ttfamily 1904.04053}}].

\bibitem{Herceg:2023pmc}
N.~Herceg, T.~Juri\'c, A.~Samsarov and I.~Smoli\'c, \emph{{Metric perturbations in noncommutative gravity}}, \href{https://doi.org/10.1007/JHEP06(2024)130}{\emph{JHEP} {\bfseries 06} (2024) 130} [\href{https://arxiv.org/abs/2310.06038}{{\ttfamily 2310.06038}}].

\bibitem{Kokkotas:1988fm}
K.~D. Kokkotas and B.~F. Schutz, \emph{{Black Hole Normal Modes: A {WKB} Approach. 3. The {Reissner-Nordstrom} Black Hole}}, \href{https://doi.org/10.1103/PhysRevD.37.3378}{\emph{Phys. Rev. D} {\bfseries 37} (1988) 3378}.

\bibitem{Leaver:1990zz}
E.~W. Leaver, \emph{{Quasinormal modes of Reissner-Nordstrom black holes}}, \href{https://doi.org/10.1103/PhysRevD.41.2986}{\emph{Phys. Rev. D} {\bfseries 41} (1990) 2986}.

\bibitem{Churilova:2020bql}
M.~S. Churilova, \emph{{Black holes in Einstein-aether theory: Quasinormal modes and time-domain evolution}}, \href{https://doi.org/10.1103/PhysRevD.102.024076}{\emph{Phys. Rev. D} {\bfseries 102} (2020) 024076} [\href{https://arxiv.org/abs/2002.03450}{{\ttfamily 2002.03450}}].

\bibitem{Cardoso:2008bp}
V.~Cardoso, A.~S. Miranda, E.~Berti, H.~Witek and V.~T. Zanchin, \emph{{Geodesic stability, Lyapunov exponents and quasinormal modes}}, \href{https://doi.org/10.1103/PhysRevD.79.064016}{\emph{Phys. Rev. D} {\bfseries 79} (2009) 064016} [\href{https://arxiv.org/abs/0812.1806}{{\ttfamily 0812.1806}}].

\bibitem{Synge:1966okc}
J.~L. Synge, \emph{{The Escape of Photons from Gravitationally Intense Stars}}, \href{https://doi.org/10.1093/mnras/131.3.463}{\emph{Mon. Not. Roy. Astron. Soc.} {\bfseries 131} (1966) 463}.

\bibitem{Luminet:1979nyg}
J.~P. Luminet, \emph{{Image of a spherical black hole with thin accretion disk}}, {\emph{Astron. Astrophys.} {\bfseries 75} (1979) 228}.

\bibitem{Stefanov:2010xz}
I.~Z. Stefanov, S.~S. Yazadjiev and G.~G. Gyulchev, \emph{{Connection between Black-Hole Quasinormal Modes and Lensing in the Strong Deflection Limit}}, \href{https://doi.org/10.1103/PhysRevLett.104.251103}{\emph{Phys. Rev. Lett.} {\bfseries 104} (2010) 251103} [\href{https://arxiv.org/abs/1003.1609}{{\ttfamily 1003.1609}}].

\bibitem{Jusufi:2019ltj}
K.~Jusufi, \emph{{Quasinormal Modes of Black Holes Surrounded by Dark Matter and Their Connection with the Shadow Radius}}, \href{https://doi.org/10.1103/PhysRevD.101.084055}{\emph{Phys. Rev. D} {\bfseries 101} (2020) 084055} [\href{https://arxiv.org/abs/1912.13320}{{\ttfamily 1912.13320}}].

\bibitem{Kumara:2024obd}
A.~N. Kumara, S.~Punacha and M.~S. Ali, \emph{{Lyapunov exponents and phase structure of Lifshitz and hyperscaling violating black holes}}, \href{https://doi.org/10.1088/1475-7516/2024/07/061}{\emph{JCAP} {\bfseries 07} (2024) 061} [\href{https://arxiv.org/abs/2401.05181}{{\ttfamily 2401.05181}}].

\bibitem{Konoplya:2020cbv}
R.~A. Konoplya and A.~F. Zinhailo, \emph{{Grey-body factors and Hawking radiation of black holes in $4D$ Einstein-Gauss-Bonnet gravity}}, \href{https://doi.org/10.1016/j.physletb.2020.135793}{\emph{Phys. Lett. B} {\bfseries 810} (2020) 135793} [\href{https://arxiv.org/abs/2004.02248}{{\ttfamily 2004.02248}}].

\bibitem{Konoplya:2020jgt}
R.~A. Konoplya, A.~F. Zinhailo and Z.~Stuchlik, \emph{{Quasinormal modes and Hawking radiation of black holes in cubic gravity}}, \href{https://doi.org/10.1103/PhysRevD.102.044023}{\emph{Phys. Rev. D} {\bfseries 102} (2020) 044023} [\href{https://arxiv.org/abs/2006.10462}{{\ttfamily 2006.10462}}].

\bibitem{Konoplya:2023ppx}
R.~A. Konoplya, \emph{{Quasinormal modes and grey-body factors of regular black holes with a scalar hair from the Effective Field Theory}}, \href{https://doi.org/10.1088/1475-7516/2023/07/001}{\emph{JCAP} {\bfseries 07} (2023) 001} [\href{https://arxiv.org/abs/2305.09187}{{\ttfamily 2305.09187}}].

\bibitem{Zhang:2006ij}
Y.~Zhang and Y.~X. Gui, \emph{{Quasinormal modes of a Schwarzschild black hole surrounded by quintessence}}, \href{https://doi.org/10.1088/0264-9381/23/22/004}{\emph{Class. Quant. Grav.} {\bfseries 23} (2006) 6141} [\href{https://arxiv.org/abs/gr-qc/0612009}{{\ttfamily gr-qc/0612009}}].

\bibitem{Konoplya:2006ar}
R.~A. Konoplya and A.~Zhidenko, \emph{{Gravitational spectrum of black holes in the Einstein-Aether theory}}, \href{https://doi.org/10.1016/j.physletb.2007.03.018}{\emph{Phys. Lett. B} {\bfseries 648} (2007) 236} [\href{https://arxiv.org/abs/hep-th/0611226}{{\ttfamily hep-th/0611226}}].

\bibitem{Ding:2018nhz}
C.~Ding, \emph{{Gravitational quasinormal modes of black holes in Einstein-aether theory}}, \href{https://doi.org/10.1016/j.nuclphysb.2018.12.005}{\emph{Nucl. Phys. B} {\bfseries 938} (2019) 736} [\href{https://arxiv.org/abs/1812.07994}{{\ttfamily 1812.07994}}].

\bibitem{Yang:2022ifo}
Y.~Yang, D.~Liu, A.~\"Ovg\"un, Z.-W. Long and Z.~Xu, \emph{{Probing hairy black holes caused by gravitational decoupling using quasinormal modes and greybody bounds}}, \href{https://doi.org/10.1103/PhysRevD.107.064042}{\emph{Phys. Rev. D} {\bfseries 107} (2023) 064042} [\href{https://arxiv.org/abs/2203.11551}{{\ttfamily 2203.11551}}].

\bibitem{Thorne:1980ru}
K.~S. Thorne, \emph{{Multipole Expansions of Gravitational Radiation}}, \href{https://doi.org/10.1103/RevModPhys.52.299}{\emph{Rev. Mod. Phys.} {\bfseries 52} (1980) 299}.

\bibitem{Vishveshwara:1970cc}
C.~V. Vishveshwara, \emph{{Stability of the schwarzschild metric}}, \href{https://doi.org/10.1103/PhysRevD.1.2870}{\emph{Phys. Rev. D} {\bfseries 1} (1970) 2870}.

\bibitem{Zerilli:1974ai}
F.~J. Zerilli, \emph{{Perturbation analysis for gravitational and electromagnetic radiation in a reissner-nordstroem geometry}}, \href{https://doi.org/10.1103/PhysRevD.9.860}{\emph{Phys. Rev. D} {\bfseries 9} (1974) 860}.

\end{thebibliography}\endgroup

\end{document}